\documentclass[aps,twocolumn,preprintnumbers,amsmath,amssymb,nofootinbib,superscriptaddress,notitlepage]{revtex4-1}

\usepackage{color}
\usepackage{epsfig}
\usepackage[utf8]{inputenc}

\begin{document}

\title{Light- and heavy-quark symmetries and the $Y(4230)$, \\ $Y(4360)$, $Y(4500)$, $Y(4620)$ and $X(4630)$ resonances}

\author{Fang-Zheng Peng}
\affiliation{School of Physics, Beihang University, Beijing 100191, China}

\author{Mao-Jun Yan}\email{yanmaojun@itp.ac.cn}
\affiliation{CAS Key Laboratory of Theoretical Physics, 
  Institute of Theoretical Physics, \\
  Chinese Academy of Sciences, Beijing 100190, China}

\author{Mario S\'anchez S\'anchez}
\affiliation{Centre d'\'Etudes Nucl\'eaires, CNRS/IN2P3, Universit\'e de Bordeaux, 33175 Gradignan, France}

\author{Manuel Pavon Valderrama}\email{mpavon@buaa.edu.cn}
\affiliation{School of Physics, Beihang University, Beijing 100191, China} 

\date{\today}


\begin{abstract} 
  \rule{0ex}{3ex}
  The heavy hadron spectrum is constrained by symmetries, of which
  two of the most important ones are heavy-quark spin and 
  SU(3)-flavor symmetries.
  Here we argue that in the molecular picture the $Y(4230)$ (or $Y(4260)$),
  the $Y(4360)$ and the recently discovered $Y(4500)$ and $Y(4620)$
  vector-like resonances are linked by these two symmetries.
  By formulating a contact-range effective field theory for
  the $D \bar{D}_1$ and $D_s \bar{D}_{s1}$ family of
  S- and P-wave charmed meson-antimeson systems, 
  we find that if the $Y(4230)$ were to be a pure $D \bar{D}_1$ molecular state,
  there would be a $D^* \bar{D}_1$ partner with a mass similar to the $Y(4360)$,
  a $D_s \bar{D}_{s1}$ partner with a mass close to the $Y(4500)$ and
  three $J=1,2$ $D_s^* \bar{D}_{s1}$ and $J=3$ $D_s^* \bar{D}_{s2}^{*}$
  bound states with a mass in the vicinity of $4630\,{\rm MeV}$,
  of which the first one ($J=1$) might correspond
  with the $Y(4620)$.
  The previous predictions can in turn be improved by modifying
  the assumptions we have used to build
  the effective field theory.
  In particular,
  if we consider the closeness of the $D^* \bar{D}_1$-$D^* \bar{D}_2^*$ and
  $D_s^* \bar{D}_{s1}$-$D_s^* \bar{D}_{s2}^*$ thresholds and include
  the related coupled channel dynamics, we predict a $J=2$
  positive C-parity state with a mass around
  $4650\,{\rm MeV}$.
  This hidden-strange and hidden-charm state might in turn be identified
  with the $X(4630)$ that has been discovered past year by the LHCb
  in the $J/\psi \phi$ invariant mass distribution.
\end{abstract}

\maketitle

\section{Introduction}

The $Y(4260)$ -- or $Y(4230)$, as later measurements have converged to a lower
value of its mass -- and $Y(4360)$ were originally observed by the BaBar
collaboration in the initial state radiation process
$e^+e^-\to \gamma_{\rm ISR}\,J/\psi \pi^+\pi^-$~\cite{Aubert:2005rm} and
in the $e^+ e^- \to \psi(2S) \pi^+ \pi^-$ invariant mass
distribution~\cite{Aubert:2006ge}, respectively.
Both are charmonium-like vector states that were not expected
in the quark-model and their nature still remains elusive.
The $Y(4230)$ has been proposed to have a sizable $D_1 \bar{D}$ molecular
component~\cite{Liu:2005ay,Ding:2008gr,Wang:2013cya,Wang:2013kra,Cleven:2013mka,Chen:2019mgp,Ji:2022blw},
which is the premise we will use in this work,
though there are other interesting conjectures about its nature~\cite{Zhu:2005hp,LlanesEstrada:2005hz,Maiani:2005pe,Dubynskiy:2008mq,MartinezTorres:2009xb,Li:2013ssa,Faccini:2014pma,Dubnicka:2020xoh}.
The $Y(4360)$ is also suspected to have non-charmonium
components~\cite{Dubynskiy:2008mq,Li:2013ssa,Durkaya:2015wra}.
A recent observation in $\eta J/\psi$ by
the BESIII collaboration~\cite{Ablikim:2020cyd}
provides a mass and width of
\begin{eqnarray}
  M_Y &=& (4218.6 \pm 3.8 \pm 2.5) \,{\rm MeV} \, , \label{eq:MY} \\
  \Gamma_Y &=& (82.0 \pm 5.7 \pm 0.4)\,{\rm MeV} \, , \label{eq:GY}
\end{eqnarray}
for the $Y(4230)$ and
\begin{eqnarray}
  M_{Y^*} &=& (4382.0 \pm 13.3 \pm 1.7) \,{\rm MeV} \, , \\
  \Gamma_{Y^*} &=& (135.9 \pm 60.8 \pm 22.5)\,{\rm MeV} \, ,
\end{eqnarray}
for the $Y(4360)$, where we notice that in~\cite{Ablikim:2020cyd}
these two resonances are referred to as $Y(4220)$ and $Y(4390)$,
following the nomenclature of their previous observation
in $\pi^+ \pi^- h_c$~\cite{BESIII:2016adj}.
However, the experimental situation is not necessarily clear with respect to
whether the $Y(4230)$ and $Y(4360)$ are each a unique states or not,
with the 2020 edition of the Review of Particle Physics
(RPP)~\cite{Zyla:2020zbs} subdividing each one into two entries,
$\psi(4230)$, $\psi(4260)$ for the $Y(4230)$ and
$\psi(4360)$, $\psi(4390)$ for the $Y(4360)$ (while the more recent
2022 edition does not~\cite{Workman:2022preview}).
Here we will treat the $Y(4230)$ and $Y(4360)$ as if each were a unique state.

It is interesting to notice the most recent observation of the $Y(4230)$
by the BESIII collaboration~\cite{BESIII:2022joj}
(in $e^{+} e^{-} \to K^{+} K^{-} J / \psi$),
which gives
\begin{eqnarray}
  M_Y &=& (4225.3 \pm 2.3 \pm 21.5) \,{\rm MeV} \, , \\
  \Gamma_Y &=& (72.9 \pm 6.1 \pm 30.8)\,{\rm MeV} \, ,
\end{eqnarray}
and finds a new resonance with a mass and width of
\begin{eqnarray}
  M_{Y_s} &=& (4484.7 \pm 13.3 \pm 24.1) \,{\rm MeV} \, , \label{eq:MYs} \\
  \Gamma_{Y_s} &=& (111.1 \pm 20.1 \pm 15.2)\,{\rm MeV} \, ,
\end{eqnarray}
which is referred to as the $Y(4500)$, in reference to its mass.
In this work we will conjecture that the $Y(4500)$ is the hidden-strange
partner of the $Y(4230)$.

Yet, there are other two hidden-charm resonances of interest to us.
The Belle collaboration has observed~\cite{Jia:2019gfe} a resonance
resembling the $Y(4230)$ in the $e^{+} e^{-} \to D_s^+ D_{s1}(2526)^- + c.c$
process, with mass and width
\begin{eqnarray}
  M_{Y_s^*} &=& (4625.9_{-6.0}^{+6.2} \pm 0.3)\,{\rm MeV} \, , \\
  \Gamma_{Y_s^*} &=& (49.8_{-11.5}^{+13.9} \pm 4.0) \, {\rm MeV} \, ,
\end{eqnarray}
which we will refer to as $Y(4620)$.
This discovery has been followed by a very similar resonance also found
by Belle~\cite{Belle:2020wtd} in the $e^{+} e^{-} \to D_s^+ D_{s2}(2573)^- + c.c$
process with compatible mass and width
\begin{eqnarray}
  M_{Y_s^*} &=& (4619.8_{-8.0}^{+8.9} \pm 2.3)\,{\rm MeV} \, , \label{eq:MYs-star}
  \\
  \Gamma_{Y_s^*} &=& (47.0_{-14.8}^{+31.3} \pm 4.6)
        {\rm MeV} \, ,
\end{eqnarray}
which we will presume to be the same state as the one found
in~\cite{Jia:2019gfe}.
Finally, the LHCb collaboration has discovered a series of
$c\bar{c} s\bar{s}$ and $c\bar{c} u\bar{s}$ resonances
in the $J/\psi \phi$ and $J/\psi K^+$ invariant
  mass distributions~\cite{Aaij:2021ivw},
of which we will highlight the $X(4630)$:
\begin{eqnarray}
  M_{X_s} &=& (4626_{-110}^{+18} \pm 16)\,{\rm MeV} \, ,\\
  \Gamma_{X_s} &=& (174_{-73}^{+134} \pm 27)
        {\rm MeV} \, .
\end{eqnarray}
This discovery connects with previous predictions of $D_{s}^* \bar{D}_{s2}^*$
virtual states with a similar mass~\cite{Dong:2021juy}, but it has also
been explained as a bound state~\cite{Yang:2021sue} and there are
non-molecular explanations too~\cite{Agaev:2022iha,Wang:2021ghk}.

Here we will explore the possible relation among these five states
within the molecular picture, where we will assume that
they are S- and P-wave charmed meson-antimesons
bound states (where, for the P-wave charmed mesons, we will be referring to
the ones with total light-quark spin $J_L = \tfrac{3}{2}$).
The tentative molecular interpretations we will propose for these five states
are listed in Table \ref{tab:thresholds}, together with the binding energies.
For describing them as molecules
we will formulate a non-relativistic effective field theory (EFT)
which incorporates SU(3)-flavor and heavy-quark spin symmetry (HQSS),
in line with previous EFT works for S-wave charmed meson-antimeson
systems~\cite{AlFiky:2005jd,Fleming:2007rp,Mehen:2011yh,Valderrama:2012jv,Nieves:2012tt,HidalgoDuque:2012pq,Baru:2016iwj}.

Before presenting the complete derivation of our formalism, we will
provide a brief, heuristic overview of our results.
If we consider the $Y(4230)$ and $Y(4360)$ as $1^{--}$ $D \bar{D}_1$ and
$D^* \bar{D}_1$ molecular states and in a first approximation ignore
the kinetic energy of the charmed mesons (as they are heavy),
their masses will be given by
\begin{eqnarray}
  m(Y(4230)) &=& m(D) + m(D_1) - \langle V_0(Y) \rangle \, , \\
  m(Y(4360)) &=& m(D^*) + m(D_1) - \langle V_0(Y^*) \rangle \, ,
\end{eqnarray}
where $\langle V_I(Y^{(*)}) \rangle$ is the expected value of the potential
binding the previous meson-antimeson systems in the isospin $I=0$ configuration.
Conversely, if the $Y(4500)$ and $Y(4620)$ happen to be $1^{--}$
$D_s \bar{D}_{s1}$ and $D_s^* \bar{D}_{s1}$ molecules, their
potentials should be directly related to those of
the $Y(4230)$ and $Y(4360)$ via HQSS and
SU(3)-flavor symmetry.
If we additionally assume that the potentials in the $I=1$ configurations
are negligible in comparison with the $I=0$ ones (e.g. if the potential
is given by vector-meson exchange), we end up with
\begin{eqnarray}
  m(Y(4500)) &=& m(D_s) + m(D_{s1}) - \frac{1}{2}\langle V_0(Y) \rangle \, ,
  \label{eq:MY4500} \\
  m(Y(4620)) &=& m(D_s^*) + m(D_{s1}) - \frac{1}{2}\langle V_0(Y^*) \rangle \, .
  \label{eq:MY4620}
\end{eqnarray}
This means that we can calculate the masses of the $Y(4500)$ and $Y(4620)$
resonances from the ones of the $Y(4230)$ and $Y(4360)$, respectively,
leading to
\begin{eqnarray}
  m(Y(4500)) &=& 4468.1 \, {\rm MeV} \, , \\
  m(Y(4620)) &=& 4623.0 \, {\rm MeV} \, , 
\end{eqnarray}
which are in line with their experimental masses,
see Eqs.~(\ref{eq:MYs}) and (\ref{eq:MYs-star}).
This suggests that the $Y(4500)$ and $Y(4620)$ are simply the hidden-strange
partners of the $Y(4230)$ and $Y(4360)$.
In contrast, the explanation of the $X(4630)$ is more involved and
we will discuss it later, as it requires to consider the explicit
HQSS structure of the potential as well as
the inclusion of coupled channel effects.

In the following lines we will explain in more detail the previous arguments and
the constraints imposed by HQSS and SU(3)-flavor symmetry to the family of
molecular states composed of an S-wave and P-wave
charmed meson-antimeson pair.
The manuscript is structured as follows: in Sect.~\ref{sec:hqss} we write
down the lowest order contact-range potential compatible with HQSS
for these systems; in Sect.~\ref{sec:su3} we consider
the effects of SU(3)-flavor; in Sect.~\ref{sec:predictions} we propose
two possible power countings by which to organize
our limited knowledge of these systems.
Finally, in Sect.~\ref{sec:conclusions} we summarize our findings.
In addition Appendix \ref{app:OPE} analyzes whether pion exchanges
are perturbative, while in Appendix \ref{app:transition-operators}
we explain a few technical details about the spin operators
appearing in these bound states
and Appendix~\ref{app:phenomenology} describes two phenomenological
models to which we compare our EFT results.

\begin{table}[!ttt]
	\begin{tabular}{|ccccc|}
	  \hline
          State & M & System & $M_{\rm th}$ & $B$ \\
          \hline
          $Y(4230)$ & $4218.6 \pm 4.5$~\cite{Ablikim:2020cyd} &
          $D_1 \bar{D}$ & $4289.4$ & $70.8 \pm 4.5$ \\
          $Y(4360)$ & $4382.0 \pm 13.4$~\cite{Ablikim:2020cyd}
          & $D_1 \bar{D}^*$ & $4430.7$ & $48.7 \pm 13.4$ \\
          $Y(4500)$ & $4484.7 \pm 27.5$~\cite{BESIII:2022joj}
          & $D_{s1} \bar{D}_s$ & $4503.5$ & $18.8 \pm 27.5$ \\
          $Y(4620)$ & $4625.9^{+6.2}_{-6.0}$~\cite{Jia:2019gfe}
          & $D_{s1} \bar{D}_s^*$ & $4647.3$
          & $52.8^{+6.0}_{-6.2}$ \\
          $X(4630)$ & $4626^{+24}_{-111}$~\cite{Aaij:2021ivw} &
          $D_{s1} \bar{D}_s^*$ & $4647.3$ & $21^{+111}_{-24}$ \\
          \hline
        \end{tabular}
        \caption{Masses of the Y-states considered in this work and
          their comparison with the closest relevant charmed
          meson-antimeson threshold.
          The column ``State'' refers to the particular state, ``M'' is
          its mass (as taken from the reference indicated in the table,
          where statistical and systematic errors have been summed
          in quadrature), ``System'' its tentative charmed
          meson-antimeson interpretation,
          ``$M_{\rm th}$'' the threshold mass of said system and ``$B$''
          its binding energy ($B = M_{\rm th} - M$).
          All masses and binding energies are expressed in ${\rm MeV}$.
	}
	\label{tab:thresholds}
\end{table}

\section{Heavy-quark spin symmetry and the contact-range potential}
\label{sec:hqss}

We begin by explaining how HQSS constrains the S-wave contact-range
interaction between the S- and P-wave charmed mesons.
The interaction (derived for the first time in Ref.~\cite{Guo:2017jvc})
contains four components and four unknown coupling constants,
which increase to eight once we consider isospin and
SU(3)-flavor symmetries.

The choice of a contact-range potential is grounded on the observation
that this is expected to be the lowest order description within
an effective field theory (EFT) with charmed mesons and
pions as degrees of
freedom~\cite{Fleming:2007rp,Mehen:2011yh,Valderrama:2012jv}.

Pions are perturbative at low energies, as we explain in detail
in Appendix~\ref{app:OPE} where we compare the sizes of
iterated and tree-level one pion exchange (OPE).
There we find that the inclusion of non-perturbative OPE effects is not
required for momenta below $800\,{\rm MeV}$ (at the minimum),
which basically is of the same order of the $\rho$ mass, i.e.
the momentum scale at which we expect the EFT description
not to be valid anymore.
It is worth noticing that this is analogous to what happens with OPE
in two-body systems composed of S-wave charmed
mesons~\cite{Fleming:2007rp,Valderrama:2012jv}, and compatible with
recent calculations of the effect of pions on the binding energy
of S- and P-wave charmed meson-antimeson molecules~\cite{Ji:2022blw},
which happens to be small.
Consequently, we will consider pions to be subleading.

In contrast, contact-range couplings are leading: the EFT description of
shallow bound states requires the promotion of this type of
couplings to leading order~\cite{vanKolck:1998bw}.
Of course there will be limitations with this description: the S- and P-wave
charmed meson-antimeson molecular candidates are not shallow,
but show binding energies for which the EFT expansion
is not expected to converge particularly well.
For instance, if we consider the $Y(4230)$ to be a $D \bar{D}_1$ bound state
then it will have a binding energy of about $B \sim 70\,{\rm MeV}$ and
a binding momentum $\gamma = \sqrt{2 \mu B} \sim 390\,{\rm MeV}$
(where $\mu$ is the reduced mass of the system).
From this we expect the expansion parameter of the EFT to be the ratio of
this binding momentum over the rho meson mass or
$\gamma / m_{\rho} \sim 0.5$.
This ratio technically lies within the range of validity of a possible EFT
description, but will result in poor convergence.

Considerations of convergence aside, the four components of the S-wave
contact-range potential can be characterized as charge, dipolar magnetic,
dipolar electric and quadrupolar magnetic in an analogy
with the multipolar expansion of electromagnetic
interactions (check, e.g. Refs.~\cite{Peng:2020xrf,Peng:2021hkr}).
We first show how this decomposition arises by writing down the meson-meson
interaction in the standard superfield notation.
Then, by noticing that the interaction only depends on the light-spin of
the heavy mesons, we will rewrite the meson-meson contact-range
potential in a more compact form using a notation that exploits
this observation.

\subsection{Heavy superfield notation}

From HQSS we expect heavy hadron interactions
to be independent of the spin of the heavy-quarks within them.
This symmetry is implemented at the practical level by writing down superfields
that group together the ground and excited states of a heavy hadron.
For the S-wave charmed mesons $D$ and $D^*$, their superfield is
\begin{eqnarray}
  H = \frac{1}{\sqrt{2}}\,
  \left[ D + \vec{\sigma} \cdot {\vec{D}\,}^* \right] \, ,
\end{eqnarray}
while for the P-wave charmed mesons $D_1$ and $D_2^*$ (that is, we are
considering the $P$-wave meson for which the total light-spin of
the light-quark is $J_L = \tfrac{3}{2}$) we have
\begin{eqnarray}
  T_i = \frac{1}{\sqrt{2}}\,\left[ D_{2,ij}^* \sigma_j +
    \sqrt{\frac{3}{2}}\, D_{1,j} ( \delta_{ij} - \frac{1}{3} \sigma_j \sigma_i)
    \right] \, ,
\end{eqnarray}
where the indices in $D_{1,j}$ and $D_{2,ij}^*$ refer to their polarization states
in the Cartesian basis, while $\sigma_i$ with $i=1,2,3$ are the Pauli matrices.
The superfields $H$ and $T_i$ defined here correspond with the non-relativistic
limit of the superfields in~\cite{Falk:1992cx}.

With the previous superfields, the most general S-wave contact-range Lagrangian
between the $H$ and $T$ fields with no derivatives is
\begin{eqnarray}
  \mathcal{L}_{\rm cont} &=&
  C_a\,{\rm Tr}\left[ H^{\dagger} H \right]\,
  {\rm Tr}\left[ T_k^{\dagger} T_k \right] \nonumber \\ &+&
  C_b\,{\rm Tr}\left[ H^{\dagger} \sigma_i H \right]\,
  {\rm Tr}\left[ T_k^{\dagger} \sigma_i T_k \right] \nonumber \\ &+&
  D_b\,{\rm Tr}\left[ H^{\dagger} T_i \right]
  {\rm Tr}\left[ T_i^{\dagger} H \right] \nonumber \\ &+&
  D_c\,{\rm Tr}\left[ H^{\dagger} T_i \sigma_j\right]
  {\rm Tr}\left[ T_i^{\dagger} H \sigma_j \right] \, ,
\end{eqnarray}
and contains four couplings (which become eight once we include isospin or
SU(3)-flavor symmetry).
For simplicity we have not indicated either whether
we are dealing with heavy mesons or antimesons.
By expanding this Lagrangian in terms of the heavy meson fields and reordering
the indices properly, we will arrive at the potential
in Table \ref{tab:molecules}.
In the particular case of the $H$ and $T$ superfields, this is painstakingly
difficult, which is why we will also derive the potential
in a second and more convenient notation.

\subsection{Light subfield notation}

Now we rewrite the interaction between an $S$-wave and $P$-wave heavy meson
within the aforementioned second notation.
In this alternative notation, which we might call light-quark notation or
light subfield notation, we only explicitly consider the spin of
the light quarks, while the heavy quarks act basically
as spectators.
In particular  we can represent the $S$- and $P$-wave mesons
by the non-relativistic light-quark subfields
\begin{eqnarray}
  q_s(J_L^P = \frac{1}{2}^+)
  \quad \mbox{and} \quad q_p(J_L^P = \frac{3}{2}^-) \, ,
\end{eqnarray}
where we indicate the spin and parity of the light-quark degrees of freedom
in parentheses.
With these fields the lowest-order contact-range Lagrangian
between an $S$-wave and $P$-wave heavy hadron is
\begin{eqnarray}
  \mathcal{L} &=&
  C_a\,q_{s}^{\dagger} q_{s}\,\,
  q_{p}^{\dagger} q_{p} +
  C_b\,\sum_i\,(q_{s}^{\dagger} {\sigma}_{Li} q_{s})\,
  (q_{p}^{\dagger} {S}_{Li} q_{p})
  \nonumber \\ &+&
  D_b\,\sum_i\,(q_{p}^{\dagger} \Sigma_{Li}^\dagger q_{s})\,
  (q_{s}^{\dagger} {\Sigma}_{Li} q_{p}) \nonumber \\
  &+&
  D_c\,\sum_{ij}\,(q_{p}^{\dagger} Q_{Lij}^{\dagger} q_{s})\,
  (q_{s}^{\dagger} Q_{Lij} q_{p})
  \, , \label{eq:LC}
\end{eqnarray}
where $C_a$, $C_b$, $D_b$ and $D_c$ are coupling constants, ${\sigma}_{Li}$
and ${S}_{Li}$ (with $i=1,2,3$) the spin operators of the light quarks inside
the S- and P-wave heavy mesons (the Pauli and
the spin-$\tfrac{3}{2}$ matrices, respectively)
and $\vec{\Sigma}_L$ a spin operator involved
in the transition from spin-$\tfrac{1}{2}$ to -$\tfrac{3}{2}$ and vice versa
(the matrix elements of which can be consulted
in Appendix \ref{app:transition-operators}),
while $Q_{Lij}$ is a tensor spin operator
that is defined as:
\begin{eqnarray}
  Q_{Lij} &=& \frac{1}{2} \left[
    \sigma_{Li} \Sigma_{Lj} + \sigma_{Lj} \Sigma_{Li} \right] \, . 
\end{eqnarray}
The most important of these couplings will be $D_b$, which we will assume
to play a very important role in the description of the $Y(4230)$.

From the previous contact-range Lagrangian we obtain the following
contact-range potential
\begin{eqnarray}
  V_C = C_a + C_b \, \vec{\sigma}_{L} \cdot \vec{S}_{L} +
  D_b\,\vec{\Sigma}_L^{\dagger} \cdot \vec{\Sigma}_L +
  D_c\,Q_L^{\dagger} \cdot Q_L \, , \nonumber \\ \label{eq:VC}
\end{eqnarray}
where we define the product of the two tensor spin operators as
\begin{eqnarray}
  Q_L^{\dagger} \cdot Q_L = \sum_{ij} Q_{Lij}^{\dagger} Q_{Lij} \, .
\end{eqnarray}
This potential is written in terms of the light-spin operators,
which do not directly correspond with the heavy-meson spin operators.
For applying the potential in the heavy-meson basis, we have to supplement
the previous potential with a series of rules for translating
the light-spin operator into the heavy-meson spin operators.
For the S-wave charmed mesons, $D$ and $D^*$, these rules are
\begin{eqnarray}
  \langle D | \vec{\sigma}_L | D \rangle &=& 0 \, , \\
  \langle D^* | \vec{\sigma}_L | D^* \rangle &=& \vec{J}_1 \, ,
\end{eqnarray}
where $\vec{J}_1$ stands for the spin-$1$ matrices
as applied to the $D^*$ meson.
For the P-wave charmed mesons, $D_1$ and $D_2^*$, the rules are
\begin{eqnarray}
  \langle D_1 | \vec{S}_L | D_1 \rangle &=& \frac{5}{4} \, \vec{S}_1 \, , \\
  \langle D_2 | \vec{S}_L | D_2 \rangle &=& \frac{3}{4} \, \vec{S}_2 \, ,
\end{eqnarray}
where $\vec{S}_1$ and $\vec{S}_2$ are the spin-$1$ and -$2$ matrices
as applied to the $D_1$ and $D_2^*$ mesons, respectively.

Finally we have the transition from the S- to P-wave mesons,
which involve the $\vec{\Sigma}_L$ and $Q_{Lij}$ operators. 
The $\vec{\Sigma}_L$ matrices are used to represent
the $D \to D_1$, $D^* \to D_1$ and $D^* \to D_2^*$ transitions,
which are generated for instance by electric-type dipole
E1 vector-meson exchange operators~\cite{Peng:2021hkr}.
The transformation rules are
\begin{eqnarray}
  \langle D | \vec{\Sigma}_L | D_1 \rangle &=&
  -\sqrt{\frac{2}{3}}\, \vec{\epsilon} \, , \label{eq:SigmaL-1} \\
  \langle D^* | \vec{\Sigma}_L | D_1 \rangle &=&
  -\frac{1}{\sqrt{6}} \, \vec{S} \, , \\
  \langle D^* | \vec{\Sigma}_L | D_2 \rangle &=&
    \vec{\Sigma} \, , \label{eq:SigmaL-3}
\end{eqnarray}
plus their hermitian conjugates when we exchange the S- and P-wave mesons,
where $\vec{\epsilon}$ is the polarization vector of the $D_1$ meson,
$\vec{S}$ the spin-1 operator and $\vec{\Sigma}$ the operator
for the spin-$1$ to spin-$2$ transition (which we write down
in Appendix \ref{app:transition-operators}).
The choice of different symbols ($\vec{J}_1$, $\vec{S}_1$ and $\vec{S}$)
for what is essentially the same spin-1 matrices is made to emphasize
that they are operating on different states, as it matters
where they are sandwiched in between.

Next, the $Q_{Lij}$ matrices represent the $D \to D_2$, $D^* \to D_1$ and
$D^* \to D_2^*$ transitions that can be generated by a magnetic-type
quadrupolar M2 vector-meson exchange operator~\cite{Peng:2021hkr}.
The transformation rules for the quadrupolar operator $Q_{Lij}$ are
\begin{eqnarray}
  \langle D | Q_{Lij} | D_1 \rangle &=& 0 \, , \label{eq:QL-1} \\
  \langle D | Q_{Lij} | D_2^* \rangle &=& -\epsilon_{tij} \, , \\
  \langle D^* | Q_{Lij} | D_1 \rangle &=&
  \sqrt{\frac{3}{2}}\,Q_{ij} \, , \\
  \langle D^* | Q_{Lij} | D_2^* \rangle &=& Q_{12ij} \label{eq:QL-4} \, ,
\end{eqnarray}
where $\epsilon_{tij}$ is the tensor polarization of the $D_2^*$ charmed meson,
$Q$ the quadrupolar spin operator for spin-1, which is defined as
\begin{eqnarray}
  Q_{ij} = \frac{1}{2} \left[ S_{i} S_{j} + S_{j} S_{i} \right] -
  \frac{1}{3} {\vec{S}\,}^2\,\delta_{ij} \, ,
\end{eqnarray}
where we remind that $\vec{S}$ refers to the spin-1 matrices when sandwiched
between a $D^*$ and a $D_1$ charmed meson, and $Q_{12ij}$ is defined as
\begin{eqnarray}
  Q_{12ij} &=& \frac{1}{2} \left[
    J_{1i} \Sigma_{j} + J_{1j} \Sigma_{i} \right] \, . 
\end{eqnarray}

\subsection{C-parity convention}

For calculating the potential in the different $J^{PC}$ configurations
it is useful to specify the C-parity convention used,
which in our case is
\begin{eqnarray}
  C | D \rangle = | \bar{D} \rangle \quad &,&
  \quad C | D^* \rangle = | \bar{D}^* \rangle \, , \\
  C | D_1 \rangle = | \bar{D}_1 \rangle \quad &,&
  \quad C | D_2^* \rangle = | \bar{D}_2^* \rangle \, ,
\end{eqnarray}
where $C$ is the operator transforming particles into antiparticles.
That is, we always use the same sign.
With this convention, a two-body meson-antimeson state with mesons $A$ and $B$
\begin{eqnarray}
  | A\bar{B}(\eta) \rangle = \frac{1}{\sqrt{2}}\,\left[ | A \bar{B} \rangle
    + \eta | B \bar{A} \rangle \right] \, ,
\end{eqnarray}
where $A$, $B$ are two different combinations of the S- and P-wave charmed
mesons considered here, will have C-parity $C = \eta (-1)^{S - S_A - S_B}$,
with $S$ the total spin of the $A\bar{B}$ system and $S_A$, $S_B$
the spins of mesons $A$ and $B$, and $\eta = \pm 1$ a sign.

\begin{table}[!ttt]
\begin{tabular}{|ccc|}
\hline\hline
  Molecule  & $J^{PC}$ & $V$ \\
  \hline
  $D \bar{D}_1$ & $1^{-+}$ & $C_a + \frac{2}{3}\,D_b$ \\
  $D \bar{D}_1$ & $1^{--}$ & $C_a - \frac{2}{3}\,D_b$ \\
  \hline
  $D \bar{D}_2^*$ & $2^{-+}$ & $C_a + D_c$ \\
  $D \bar{D}_2^*$ & $2^{--}$ & $C_a - D_c$ \\
  \hline
  $D^* \bar{D}_1$ & $0^{-+}$ & $C_a - \frac{5}{2}\,C_b - \frac{1}{3}\,D_b + \frac{5}{2}\,D_c$ \\
  $D^* \bar{D}_1$ & $0^{--}$ & $C_a - \frac{5}{2}\,C_b + \frac{1}{3}\,D_b - \frac{5}{2}\,D_c$ \\
  $D^* \bar{D}_1$ & $1^{-+}$ & $C_a - \frac{5}{4}\,C_b + \frac{1}{6}\,D_b + \frac{5}{4}\,D_c$ \\
  $D^* \bar{D}_1$ & $1^{--}$ & $C_a - \frac{5}{4}\,C_b - \frac{1}{6}\,D_b - \frac{5}{4}\,D_c$ \\
  $D^* \bar{D}_1$ & $2^{-+}$ & $C_a + \frac{5}{4}\,C_b + \frac{1}{6}\,D_b + \frac{1}{4}\,D_c$ \\
  $D^* \bar{D}_1$ & $2^{--}$ & $C_a + \frac{5}{4}\,C_b - \frac{1}{6}\,D_b - \frac{1}{4}\,D_c$ \\
  \hline
  $D^* \bar{D}_2^*$ & $1^{-+}$ & $C_a - \frac{9}{4}\,C_b + \frac{1}{6}\,D_b - \frac{3}{4}\,D_c$ \\
  $D^* \bar{D}_2^*$ & $1^{--}$ & $C_a - \frac{9}{4}\,C_b - \frac{1}{6}\,D_b + \frac{3}{4}\,D_c$ \\
  $D^* \bar{D}_2^*$ & $2^{-+}$ & $C_a - \frac{3}{4}\,C_b - \frac{1}{2}\,D_b + \frac{5}{4}\,D_c$ \\
  $D^* \bar{D}_2^*$ & $2^{--}$ & $C_a - \frac{3}{4}\,C_b + \frac{1}{2}\,D_b - \frac{5}{4}\,D_c$ \\
  $D^* \bar{D}_2^*$ & $3^{-+}$ & $C_a + \frac{3}{2}\,C_b + \,D_b + \frac{1}{2}\,D_c$  \\
  $D^* \bar{D}_2^*$ & $3^{--}$ & $C_a + \frac{3}{2}\,C_b - \,D_b - \frac{1}{2}\,D_c$ \\
  \hline\hline 
\end{tabular}
\caption{Structure of the contact-range potential between
  a $D$/$D^*$ S-wave charmed meson and a $D_1$/$D_2^*$ P-wave charmed meson.
  This potential depends on four coupling constants:
  $C_a$, $C_b$, $D_b$ and $D_c$.
  From the saturation arguments of Refs.~\cite{Peng:2020xrf,Peng:2021hkr},
  for isoscalar molecules we expect: $C_a < 0$, $C_b < 0$,
  $D_b > 0$ and $D_c > 0$ (check also the discussion
  in Sect.~\ref{subsec:proposing}).
  If this argument is correct, there will be a series of configurations
  that are guaranteed to be attractive: $1^{--}$ $D\bar{D}_1$,
  $2^{--}$ $D\bar{D}_2^*$, $2^{--}$ $D^* \bar{D}_1$
  and $3^{--}$ $D^* \bar{D}_2^*$.
}
\label{tab:molecules}
\end{table}

\section{Light SU(3)-flavor symmetry}
\label{sec:su3}

The light-flavor structure of the contact-range interaction potentially
provides further information about heavy meson-antimeson molecules.
The constraints that SU(3)-flavor symmetry imposes happen to be really simple:
as the charmed mesons (antimesons) belong to the $\bar{3}$ ($3$)
representation of SU(3)-flavor, their molecules will be the
sum of a singlet and octet representation
($3 \otimes \bar{3} = 1 \oplus 8$).
That is, the contact-range potential can be decomposed as a sum of
a singlet and octet component, i.e.
\begin{eqnarray}
  V_C = \lambda_S\, V_C^{(S)} + \lambda_O\, V_C^{(O)} \, , 
\end{eqnarray}
where the specific linear combination depends on the isospin and strange
content of the molecule.
In particular we have
\begin{eqnarray}
  V_C(q\bar{q}, I=0) &=&
  \frac{2}{3} V_C^{(S)} + \frac{1}{3} V_C^{(O)} \, ,  \\
  V_C(q\bar{q}, I=1) &=& V_C^{(O)} \, ,  \\
  V_C(s\bar{s}, I=0) &=&
 \frac{1}{3} V_C^{(S)} + \frac{2}{3} V_C^{(O)} \, , 
\end{eqnarray}
where $q\bar{q}$ and $s\bar{s}$ refer to the light-quark content of the
charmed meson-antimeson system under consideration, with
the meaning of $q$ here restricted to $q = u,d$.

Here two comments are in order: first, the strange and non-strange $I=0$
configurations can mix, which means for instance that the $Y(4230)$ and
$Y(4360)$ will have a strange component.
This type of coupled-channel effect is expected to be small though and
we will ignore it.
Second, it is worth noticing that the potential in the $s\bar{s}$
configuration is the average of the isoscalar and isovector ones
\begin{eqnarray}
  V_C(s\bar{s}, I=0) &=&
  \frac{1}{2} \, V_C(q\bar{q}, I=0) + \frac{1}{2} \, V_C(q\bar{q}, I=1) \, .
  \nonumber \\
\end{eqnarray}
This happens to be important because from phenomenological models
we expect the potential in the isovector configuration to be
considerably weaker than in the isoscalar one.
The reason is that vector meson exchange cancels out in the $I=1$ channel
and thus the interaction in this channel depends either on SU(3)-flavor
symmetry breaking effects or exchange of more massive mesons
(two possibilities being vector charmonia~\cite{Aceti:2014kja,Aceti:2014uea,He:2015mja,Dong:2021juy} or axial meson exchange~\cite{Yan:2021tcp}).
Naively, we expect these contributions to be weaker
than vector meson exchange.

\section{Predictions and power countings}
\label{sec:predictions}

When making predictions with the contact-range theory we have formulated,
there will be a significant problem:
the lowest-order contact-range potential contains a total of eight unknown
couplings (or four if we concentrate in one spin/flavor sector only).
In principle the determination of these couplings requires the identification of
eight molecular candidates containing different linear combinations
of the couplings.
Owing to a lack of a suitable number of molecular candidates,
this is not a feasible prospect at the moment.

There is a way out of this predicament though: we do not expect
all the couplings to be equally important and a few couplings
might provide a larger contribution to binding than others.
This idea might have support in phenomenology: for instance,
vector meson exchange together with vector meson dominance~\cite{Peng:2021hkr}
suggest a really strong electric dipole term (and to a lesser extent,
magnetic quadrupole term)
mediating the $D^{(*)} \to D_{1(2)}^{(*)}$ family of transitions.
This type of vector meson exchange generates a potential with the same
operator structure as the $D_b$ term (or the $D_c$ term for the magnetic
quadrupole transition) in the S-wave contact-range potential.
It happens that the $D_b$ coupling might be strong enough as to bind
the $D \bar{D}_1$ system by itself, and the same might be true
for the $D_c$ coupling and the $D^* \bar{D}_1$ system.
If this is the case, it might be possible to ignore the other couplings
at lowest order.

Within the EFT language we do not necessarily worry about the microscopic
origin of the contact-range potential we use, be it either light-meson
  exchange or shorter-range $c\bar{c}$ components of
  the wave function~\cite{Li:2009zu}.
Instead we simply point out that there are several possible
{\it power countings}, i.e. different ways in which to order
the operators of the EFT from more to less relevant.
The power counting can be simply assumed, where its suitability and correctness
can be later judged on the basis of the predictions a particular
power counting entails.
In this regard we will explore power countings in which
the $D_b$ and $D_c$ couplings are conjectured to be more relevant
than the $C_a$ and $C_b$ couplings.

\subsection{What we know and what we do not know: proposing a power counting}
\label{subsec:proposing}

In principle we have eight independent coupling constants, namely
\begin{eqnarray}
  C_{Ia} \quad , \quad C_{Ib} \quad , \quad D_{Ib} \quad \mbox{and} \quad D_{Ic}
  \, , \nonumber
\end{eqnarray}
with isospin $I = 0,1$.
However, as previously mentioned, there are not enough molecular candidates
involving S- and P-wave charmed mesons.
The most solid candidate is the $Y(4230)$, theorized to be a $1^{--}$
$D\bar{D}_1$ bound state~\cite{Liu:2005ay,Ding:2008gr,Wang:2013cya,Wang:2013kra,Cleven:2013mka,Chen:2019mgp,Ji:2022blw} (though its nature is far from clear
and other explanations
exist~\cite{Zhu:2005hp,LlanesEstrada:2005hz,Maiani:2005pe,Dubynskiy:2008mq,MartinezTorres:2009xb,Li:2013ssa,Faccini:2014pma,Dubnicka:2020xoh}).
With this we can single out one coupling or combination of couplings that
is responsible for its binding, and set the other couplings to zero.
In terms of power counting this is justified by identifying the chosen coupling
as leading order, while all the other couplings will be subleading.

For choosing the coupling, we will first note that the $D\bar{D}_1$
potential in the $J^{PC} = 1^{--}$ channel contains two couplings
\begin{eqnarray}
  V_C(D\bar{D}_1, 1^{--}) &=& C_{0a} - \frac{2}{3}\,D_{0b} \, , 
\end{eqnarray}
where this observation can be combined with phenomenological information.
In particular, the S-wave light-meson exchange potential between the S- and
P-wave charmed mesons can be written as
\begin{eqnarray}
  V = V_a + V_b \, \vec{\sigma}_{L} \cdot \vec{S}_{L} +
  W_b\,\vec{\Sigma}_L^{\dagger} \cdot \vec{\Sigma}_L +
  W_c\,Q_L^{\dagger} \cdot Q_L \, , \nonumber \\
  \label{eq:V-components}
\end{eqnarray}
which at high momenta will receive contributions
from the sigma, rho and omega mesons.
The explicit form of these components was calculated in~\cite{Peng:2021hkr},
yielding
\begin{eqnarray}
  V_a(\vec{q}\,) &=& - \frac{g_{S1} g_{S2}}{m_S^2 + {\vec{q}\,}^2} 
  + \left(\vec{\tau}_1 \cdot \vec{\tau}_2 + \zeta \right)\,
  \frac{g_{V1} g_{V2}}{m_V^2 + {\vec{q}\,}^2} \, , \label{eq:Va} \\
  V_b(\vec{q}\,) &=& + \left( \vec{\tau}_1 \cdot \vec{\tau}_2 + \zeta \right)\,
  \frac{2}{3}\,\frac{f_{V1} f_{V2}}{6 M^2} \,
  \frac{m_V^2}{m_V^2 + {\vec{q}\,}^2} + \dots \, , \nonumber \\ \label{eq:Vb} \\
  W_b(\vec{q}\,) &=& - \left( \vec{\tau}_1 \cdot \vec{\tau}_2 + \zeta \right)\,
  \frac{{f_{V}'}^2}{4 M^2}\,
  \frac{\omega_V^2 + \frac{1}{3} \mu_V^2}{\mu_V^2 + {\vec{q}\,}^2} + \dots \, ,
  \label{eq:Wb} \\
  W_c(\vec{q}\,) &=& - \left( \vec{\tau}_1 \cdot \vec{\tau}_2 + \zeta \right)\,
  \frac{{h_{V}'}^2}{16 M^4}\,\frac{1}{5}\,
  \frac{\mu_V^4}{\mu_V^2 + {\vec{q}\,}^2} + \dots \, , \label{eq:Wc}
\end{eqnarray}
where they are written in p-space, with $\vec{q}$ the exchanged momentum.
In the expressions above $g_{Si}$, $g_{Vi}$ and $f_{Vi}$ (with $i=1,2$) are
the scalar meson and vector meson charge-like and magnetic-like couplings
for each of the heavy mesons, $f_V'$ and $h_V'$ the electric dipole and
magnetic quadrupole couplings of the vector mesons,
$M$ a scaling mass (for which a typical choice is the nucleon mass,
$M = m_N = 940\,{\rm MeV}$),
$m_S$ and $m_V$ the masses of the scalar and vector mesons and
$\mu_V^2 = m_V^2 - \omega_V^2$ the effective mass for the vector meson,
with $\omega_V = m(D_{1(2)}^*) - m(D^{(*)})$ the mass difference of
the S- and P-wave charmed mesons for the non-diagonal
(i.e. $W_b$ and $W_c$) components of the potential.
Owing to $\mu_V < m_V$, the range of the non-diagonal $W_b$ and $W_c$ components
will be larger than that of the the diagonal ones, $V_a$ and $V_b$.
Also, the $V_b$ component differs by a factor of $2/3$ to that
of Ref.~\cite{Peng:2021hkr} as a consequence of
the different normalization used
for the spin operators.

At this point it is interesting to notice that
the combination of the isospin factor $\vec{\tau}_1 \cdot \vec{\tau}_2$ and
the sign of the omega contribution ($\zeta = +1$ for a heavy meson-meson pair
and $\zeta = -1$ for the meson-antimeson case) implies that
the vector-meson exchange contributions cancel for $I=1$,
leading to the conjecture that the potential in the hidden-strange sector
is about half that of the isoscalar sector (if we assume that the scalar
meson contribution is small), e.g. Eqs.~(\ref{eq:MY4500}) and (\ref{eq:MY4620}).

Here, before discussing the size of each component, it is important to comment
on the dots in Eqs.~(\ref{eq:Va}-\ref{eq:Wc}),
which we use to indicate Dirac-delta terms that appear
as a consequence of the derivative nature of the higher-multipole
vector-meson interactions.
Following~\cite{Peng:2020xrf} we notice that the range of these delta
contributions is shorter than that of proper vector meson exchange,
from which we do not expect them to contribute significantly
to the saturation of the $C_b$, $D_b$ and $D_c$ couplings.
This observation coincides with the findings of phenomenological studies of
vector meson exchange in the charmed antimeson and charmed baryon
systems --- the pentaquarks ---, which usually require to neglect
the Dirac-delta contributions for the correct reproduction of
the known pentaquark spectrum~~\cite{Liu:2019zvb,Yalikun:2021bfm}.
Ref.~\cite{Liu:2019zvb} points out how these contributions excessively
distort the one pion exchange potential at relatively long distances,
while Ref.~\cite{Yalikun:2021bfm} suggests that a possible rationale
for neglecting the Dirac-deltas might be the exchange of
heavier light mesons not explicitly included.

Going back to the relative importance of the components of the potential,
it happens that $W_b$, which will contribute to the $D_b$
coupling, is expected to be remarkably strong,
where in Ref.~\cite{Peng:2021hkr} it was found to generate a large
contribution to the binding energy of the $Y(4230)$.
Owing to the importance of the $W_b$ component to generate the $Y(4230)$,
at low momenta we will propose the hierarchy
\begin{eqnarray}
  | W_b | >  |V_a|\, , \, |V_b| \, , \, | W_c | \, , \label{eq:hierarchy-1}
\end{eqnarray}
which might be representative of the expected hierarchy 
for the contact-range couplings $D_b$ and $C_a$, $C_b$, $D_c$.
It is worth noticing that the $W_c$ component has also been argued
to receive a strong contribution from vector meson exchange, and
that by calibrating its strength it is possible to reproduce
the mass of the $Y(4360)$~\cite{Peng:2021hkr}.
Thus, by repeating the previous arguments we arrive at
\begin{eqnarray}
  | W_b | >  | W_c | > |V_a|\, , \, |V_b|  \, , \label{eq:hierarchy-2}
\end{eqnarray}
which might indicate a strong $D_c$ coupling too.
We warn that these hierarchies are merely qualitative in nature though, and
that they are dependent on the assumption of which components of
the potential generate binding for a given molecule.
In particular, other orderings are possible.
In a first approximation we will consider the first hierarchy we have proposed,
i.e. the one given by Eq.~(\ref{eq:hierarchy-1}) in which
the $W_b$ / $D_b$ is the largest component of
the potential.

  Finally it is important to stress that besides light-meson exchanges,
  the EFT couplings will also receive contributions from
  the shorter-range $c\bar{c}$ components of
  the wave function of the $Y$ states~\cite{Li:2009zu}.
  Owing to their shorter range, these contributions are naively expected to
  have a smaller impact on the value of the EFT couplings,
  which is why we do not discuss them directly.
  If we model the attraction coming from the $c\bar{c}$ component
  as charmonium exchange,
  as proposed in~\cite{Aceti:2014kja,Aceti:2014uea,He:2015mja,Dong:2021juy},
  the size of their contributions
  to the contact-range couplings will scale as $1/m_{J/\psi}^2$ to be compared
  with $1/m_S^2$ and $1/m_V^2$ for scalar and vector mesons.
  Yet it is important to stress that within the EFT framework
  the specific origin of the contact-range couplings is inconsequential:
  they are determined from the same experimental information regardless of
  whether they come from light-meson exchanges or a compact charmonium core.

\subsection{The $Y(4230)$ and the $Y$-counting}

From the molecular interpretation of the $Y(4230)$ and the phenomenology
we have previously discussed (derived from~\cite{Peng:2021hkr}),
we propose the following power counting
\begin{eqnarray}
  D_{0b} \sim Q^{-1} \quad \mbox{and} \quad C_{0a} \, \dots \, D_{1c} \sim Q^0 \, .
\end{eqnarray}
We will call it $Y$-counting, as it is constructed to reproduce the $Y(4230)$
with a single parameter: $D_{0b}$.
Indeed, in this counting the leading order $\left({\rm LO}\right)$  potential reads
\begin{eqnarray}
  V_C^{\rm LO(Y)}(D\bar{D}_1, 1^{--}) &=& - \frac{2}{3}\,D_{0b} \, .
  \label{eq:V-LO-Y}
\end{eqnarray}
We can determine the $D_{0b}$ coupling
from solving the bound state equation
\begin{eqnarray}
  \phi(k) + \int \frac{d^3 \vec{p}}{(2\pi)^3}\,\langle k | V^{\rm LO}_C | p \rangle\,
  \frac{\phi(p)}{M_{\rm th} + \frac{p^2}{2\mu} - M_{\rm mol}} = 0 \, ,
  \nonumber \\ \label{eq:BSE}
\end{eqnarray}
where $\phi(k)$ is the vertex function, and $M_{\rm mol}$ the mass of
the molecular state we are predicting.
However, the contact-range potential is singular and has to be regularized
(and renormalized, as explained in the next paragraph),
which we do by including a separable regulator function
\begin{eqnarray}
  \langle k | V^{\rm LO}_C | p \rangle &=&  \langle k | V^{\rm LO}_C | p \rangle\,
  f(\frac{k}{\Lambda}) f(\frac{p}{\Lambda}) \, , \nonumber \\
  &=& C^{\rm LO}(\Lambda)\,f(\frac{k}{\Lambda}) f(\frac{p}{\Lambda}) \, ,
  \label{eq:VC-reg}
\end{eqnarray}
where we will choose a Gaussian regulator $f(x) = e^{-x^2}$,
while $V_C^{\rm LO}$ is the unregularized ${\rm LO}$ potential,
which is given by Eq.~(\ref{eq:V-LO-Y}) within the $Y$-counting.

  The ${\rm LO}$ potential reduces to a coupling constant that can be
  expressed as a linear combination of the subset of $C_{Ia}$, $C_{Ib}$,
  $D_{Ib}$ and $D_{Ic}$ couplings that are non-zero at ${\rm LO}$.
  We denote this coupling as $C^{\rm LO}(\Lambda)$ in the second line of
  Eq.~(\ref{eq:VC-reg}).
  It happens that for the Gaussian regulator we have chosen, the bound
  state equation (Eq.~(\ref{eq:BSE})) can be solved analytically,
  yielding:
  \begin{eqnarray}
    1 + \frac{\mu}{4 \pi^2} \,C^{\rm LO}(\Lambda) I(\gamma, \Lambda) = 0 \, ,
    \label{eq:BSE-reg}
  \end{eqnarray}
  with $\gamma = \sqrt{2 \mu (M_{\rm th }- M_{\rm mol})}$ the wave number of
  the bound state, where the loop integral $I(\gamma, \Lambda)$ reads
  \begin{equation}
    I(\gamma, \Lambda) = \sqrt{2 \pi}\,\Lambda -
    2\, e^{2 \gamma^2 / \Lambda^2}\,\pi \gamma \,
    {\rm erfc}\left( \frac{\sqrt{2} \gamma}{\Lambda} \right) \, ,
  \end{equation}
  with ${\rm erfc}\,(x)$ the complementary error function.
  In this analytic expression it can be appreciated that $I(\gamma, \Lambda)$
  depends on the sign of $\gamma$: for $\gamma > 0$ we will talk about bound
  states, which would correspond to the numerical solutions of
  Eq.~(\ref{eq:BSE}) or to negative energy solutions of
  the Schr\"odinger equation with the asymptotic behavior
  $\Psi(\vec{r}) \propto e^{-\gamma r} / r$ for $r \to \infty$.
  In contrast, for $\gamma < 0$ we will have virtual states instead, i.e.
  negative energy solutions of the Schr\"odinger equation with the asymptotic
  behavior $\Psi(\vec{r}) \propto e^{+\gamma r} / r$ for $r \to \infty$.
  If $\gamma \to 0$ both the bound and virtual state solutions lead
  to a two-body cross section of $\sigma \to 4 \pi / \gamma^2$, which
  implies that virtual states (despite not being bound) might still
  be detectable provided they are close enough to threshold.
  For this reason we will also include virtual states in our predictions.

  For renormalizing the potential, the first step is to make the couplings
  in $V_C^{\rm LO}$ dependent on the cutoff $\Lambda$, i.e. we will
  have $C^{\rm LO} = C^{\rm LO}(\Lambda)$ as already indicated
  in Eqs.~(\ref{eq:VC-reg}) and (\ref{eq:BSE-reg}).
  Alternatively, if we expand the potential in its components
  (Eq.~(\ref{eq:V-LO-Y})) what we will have is $C_{Ia} = C_{Ia}(\Lambda)$,
  $C_{Ib} = C_{Ib}(\Lambda)$, etc., that is, all couplings
  (or the particular subset that survives at ${\rm LO}$ in the power counting
  we are using, e.g. $D_{0b} = D_{0b}(\Lambda)$ for the $Y$-counting)
  will depend on the cutoff.
  The second step is to determine the couplings from the condition of
  reproducing a series of observable quantities (for instance,
  the physical masses of a candidate molecular state)
  for a given cutoff $\Lambda$.
  From this procedure we expect the relative cutoff dependence of
  the observables predicted in the EFT to scale as
  $\mathcal{O}(Q/\Lambda)$ at ${\rm LO}$.
  As a consequence by taking $\Lambda \sim M$ the cutoff dependence of
  the ${\rm LO}$ calculation will be of the order of the subleading
  order corrections. Alternatively, by varying the cutoff around
  the breakdown scale $M$ it will be possible to estimate
  the uncertainty of the ${\rm LO}$ results, though in general
  this method is known to underestimate the errors and we will
  complement it with a second method for calculating
  the errors.

In our case, we determine (or renormalize) $D_{0b}$
from by reproducing the $Y(4230)$ mass in Eq.~(\ref{eq:BSE}),
for which we will use the central value in Eq.~(\ref{eq:MY}),
i.e. $M_Y = 4218.6\,{\rm MeV}$ (the measurement of Ref.~\cite{Ablikim:2020cyd}).
We will use the cutoff $\Lambda = 0.75 \,{\rm GeV}$, which is around the rho
meson mass.
For estimating the uncertainties of our predictions,
we will first allow the cutoff to move in the $(0.5-1.0)\,{\rm GeV}$
range, which are values very commonly used in the literature.
This choice results in a coupling of
\begin{eqnarray}
  D_{0b} = + 3.47\,(7.98-2.05)\,{\rm fm}^2 \, ,
\end{eqnarray}
for $\Lambda = 0.75\,(0.5-1.0)\,{\rm GeV}$.
  Yet, we will consider also the EFT uncertainties coming from the wave
  number of the $Y(4230)$ (from our input), as well as the wave
  number of the bound states we are predicting (from our output).
  We will estimate the relative size of this error to approach
  $\gamma / m_{\rho}$ when $\gamma$ is small,
  with $\gamma = \sqrt{2\mu (M_{\rm th} - M_{\rm mol})}$
  the wave number of a given state and $m_{\rho}$ the rho meson mass.
  The final uncertainty in the two-body binding energy $B$ of the states
  we predict will be given by
\begin{eqnarray}
  && \Delta B = \nonumber \\ && \quad
  {\rm max}\,
  \bigg\{ B(\Lambda \pm \Delta \Lambda) - B(\Lambda),
  \nonumber \\ && \qquad \qquad
  B(\Lambda) \,
  \left[ {\left( 1 + \frac{\gamma_{{\rm max}}}{m_{\rho}} \right)}^{\pm 1} - 1 \right]
  \bigg\} \, , \label{eq:B-err}
\end{eqnarray}
where the first and second errors are the cutoff and intrinsic
$\mathcal{O}(Q/M)$ uncertainty, where $B = M_{\rm th} - M_{\rm p} = \frac{\gamma^2}{2 \mu}$ and $\gamma_{\rm max} = {\rm max}(\gamma, \gamma_Y)$,
with $\gamma_Y$ the wave number of the input state,
i.e. the $Y(4230)$.
The specific form of the intrinsic error converges to the more common relative
$\gamma / m_{\rho}$ error and its equivalent to it within the ${\rm LO}$
approximation
\begin{eqnarray}
  \left[ {\left( 1 + \frac{\gamma_{{\rm max}}}{m_{\rho}} \right)}^{\pm 1} - 1 \right]
  = \pm \left( \frac{\gamma_{{\rm max}}}{m_{\rho}} \right) +
  \mathcal{O}\left[ {\left(\frac{Q}{M}\right)}^2 \right] \, . \nonumber \\
  \label{eq:B-intrinsic}
\end{eqnarray}
Yet, in situations where the convergence of the EFT is slow (as it is probably
the case here) this choice is more suitable than
the usual $\gamma / m_{\rho}$ ratio.
The reason lies in the bound states for which the binding energy is close to
the limit of convergence of the EFT: as the $\gamma / m_{\rho}$ ratio
increases for a given state, the error will eventually be compatible
with the state not binding, which curiously would have not happen
if the state was less bound in the first place.
Our choice avoids this problem, while acknowledging at the same time
that there will be large uncertainties in the mass of such a state.

With $D_{0b}$ we can now predict the full spectrum of charmed meson-antimeson
molecules in the $Y$-counting.
For this, we first determine the unregularized potential for each molecular
configuration by adapting Table \ref{tab:molecules} to the $Y$-counting,
where the resulting potentials are listed
in Table~\ref{tab:Y-counting}.
We then regularize the potential with a Gaussian regulator and plug it
into the bound state equation (Eq.~(\ref{eq:BSE})).
For the masses of the S- and P-wave mesons, which are required for calculating
$M_{\rm th}$ and $\mu$ in Eq.~(\ref{eq:BSE}), we have used the isospin average of
the masses listed in the RPP~\cite{Zyla:2020zbs}.
The poles in the first (second) Riemann sheet correspond to the bound (virtual)
states predictions, which we list in Table~\ref{tab:Y-counting}.

\begin{table*}[!ttt]
  \begin{tabular}{|cccccc||cccccc|}
\hline\hline
  Molecule  & $I^G$ & $J^{PC}$ & $V^{\rm LO(Y)}$ & B & M &
  Molecule  & $I^G$ & $J^{PC}$ & $V^{\rm LO(Y)}$ & B & M \\
  \hline
  $D \bar{D}_1$ & & $1^{-+}$ & $+\frac{2}{3}\,D_{0b}$ & - & - &
  $D_s \bar{D}_{s1}$ & & $1^{-+}$ & $+\frac{2}{6}\,D_{0b}$ & - &  -\\
  $D \bar{D}_1$ & & $1^{--}$ & $-\frac{2}{3}\,D_{0b}$ &
  Input ($70.7$) & Input ($4218.6$) & 
  $D_s \bar{D}_{s1}$ & & $1^{--}$ & $-\frac{2}{6}\,D_{0b}$ &
  $9.2^{+8.4}_{-5.9}$ & $4494.2^{+5.9}_{-8.4}$
  \\
  \hline
  $D \bar{D}_2^*$ & & $2^{-+}$ & $0$ & - & - &
  $D_s \bar{D}_{s2}^*$ & & $2^{-+}$ & $0$ & - & - \\
  $D \bar{D}_2^*$ & & $2^{--}$ & $0$ & - & - &
  $D_s \bar{D}_{s2}^*$ & & $2^{--}$ & $0$ & - & - \\
  \hline
  $D^* \bar{D}_1$ & &  $0^{-+}$ & $-\frac{1}{3}\,D_{0b}$ &
  $9.0^{+8.7}_{-5.9}$ & $4421.7^{+5.9}_{-8.7}$
  &
  $D_s^* \bar{D}_{s1}$  & & $0^{-+}$ & $-\frac{1}{6}\,D_{0b}$ &
  ${(1.5_{-1.5}^{+10.6})}^V$ & ${(4645.8^{+1.5}_{-10.6})}^V$
  \\
  $D^* \bar{D}_1$ & &  $0^{--}$ & $+ \frac{1}{3}\,D_{0b}$ & - & - &
  $D_s^* \bar{D}_{s1}$ & & $0^{--}$ & $+ \frac{1}{6}\,D_{0b}$ & - & - \\
  $D^* \bar{D}_1$ & &  $1^{-+}$ & $+ \frac{1}{6}\,D_{0b}$ & - & - &
  $D_s^* \bar{D}_{s1}$  & & $1^{-+}$ & $+ \frac{1}{12}\,D_{0b}$ & - & - \\
  $D^* \bar{D}_1$ & &  $1^{--}$ & $- \frac{1}{6}\,D_{0b}$ &
  ${(2.4_{-1.9}^{+13.0})}^V$ & {${(4428.2^{+1.9}_{-13.0})}^V$} 
  &
  $D_s^* \bar{D}_{s1}$ & &  $1^{--}$ & $- \frac{1}{12}\,D_{0b}$ & 
  ${(25^{+43}_{-22})}^V$ & ${(4622^{+22}_{-43})}^V$
  \\
  $D^* \bar{D}_1$ & &  $2^{-+}$ & $+ \frac{1}{6}\,D_{0b}$ & - & - &
  $D_s^* \bar{D}_{s1}$ & &  $2^{-+}$ & $+ \frac{1}{12}\,D_{0b}$ & - & - \\
  $D^* \bar{D}_1$ & &  $2^{--}$ & $- \frac{1}{6}\,D_{0b}$
  & ${(2.4_{-1.9}^{+13.0})}^V$ & {${(4428.2^{+1.9}_{-13.0})}^V$}
  &
  $D_s^* \bar{D}_{s1}$ & &  $2^{--}$ & $- \frac{1}{12}\,D_{0b}$ &
  ${(25^{+43}_{-22})}^V$ & ${(4622^{+22}_{-43})}^V$
  \\
  \hline
  $D^* \bar{D}_2^*$ & &  $1^{-+}$ & $+ \frac{1}{6}\,D_{0b}$ & - & - &
  $D_s^* \bar{D}_{s2}^*$  & & $1^{-+}$ & $+ \frac{1}{12}\,D_{0b}$ & - & - \\
  $D^* \bar{D}_2^*$ & &  $1^{--}$ & $- \frac{1}{6}\,D_{0b}$ &
  ${(2.2_{-2.2}^{+12.9})}^V$ & ${(4467.4^{+2.2}_{-12.9})}^V$
  &
  $D_s^* \bar{D}_{s2}^*$  & & $1^{--}$ & $- \frac{1}{12}\,D_{0b}$ &
  ${(25^{+42}_{-21})}^V$ & ${(4657^{+21}_{-42})}^V$
  \\
  $D^* \bar{D}_2^*$ & &  $2^{-+}$ & $ - \frac{1}{2}\,D_{0b}$ &
  $38^{+19}_{-13}$ & $4432^{+13}_{-19}$
  &
  $D_s^* \bar{D}_{s2}^*$  & & $2^{-+}$ & $ - \frac{1}{4}\,D_{0b}$
  & $1.8_{-1.8}^{+6.6}$ & $4679.5^{+1.8}_{-6.6}$
  \\
  $D^* \bar{D}_2^*$ & &  $2^{--}$ & $+ \frac{1}{2}\,D_{0b}$ & - & - &
  $D_s^* \bar{D}_{s2}^*$  & & $2^{--}$ & $ + \frac{1}{4}\,D_{0b}$ & - & -\\
  $D^* \bar{D}_2^*$ & &  $3^{-+}$ & $+ D_{0b}$ & - & - &
  $D_s^* \bar{D}_{s2}^*$  & & $3^{-+}$ & $+ \frac{1}{2}\,D_{0b}$ & - & - \\
  $D^* \bar{D}_2^*$ & &  $3^{--}$ & $- D_{0b}$
  & $158^{+121}_{-68}$ & $4311^{+68}_{-121}$ 
  &
  $D_s^* \bar{D}_{s2}^*$ & & $3^{--}$ & $- \frac{1}{2}\,D_{0b}$ &
  $41^{+21}_{-14}$ & $4640^{+14}_{-21}$ 
  \\
  \hline\hline 
\end{tabular}
  \caption{
    Contact-range potential and bound state spectrum in the $Y$-counting.
    This counting assumes that the $Y(4230)$ is a $J^{PC} = 1^{--}$
    $D \bar{D}_1$ bound state generated by the strength of
    the $D_{b0}$ coupling.
    Meanwhile, all other couplings are considered as subleading corrections.
    The central predictions correspond to the cutoff
    $\Lambda = 0.75\,{\rm GeV}$, while the error is the maximum
    of either the cutoff dependence in the $(0.5-1.0)\,{\rm GeV}$ window or
    the intrinsic EFT truncation error, see Eq.~(\ref{eq:B-err}) and
    the related discussion for details.
    The superscript $V$ indicates a virtual state solution
    (instead of a bound state one).
    The binding energies and masses are in units of ${\rm MeV}$.
  }
\label{tab:Y-counting}
\end{table*}

\subsection{Interpretation of the $Y(4500)$ and $Y(4620)$ resonances}

If we take into consideration the ${\rm LO}$ EFT potential in the $Y$-counting
in more detail, we quickly realize that the $1^{--}$ $D_{s} \bar{D}_{s1}$ and
$0^{-+}$, $1^{--}$ and $2^{--}$ $D_{s}^* \bar{D}_{s1}$ and
$3^{--}$ $D_{s}^* \bar{D}_{s2}^*$ configurations
are attractive.
For the $1^{--}$ $D_{s} \bar{D}_{s1}$ system, we find a bound state at
\begin{eqnarray}
  M(D_{s1} \bar{D}_s, 1^{--}) &=& 4494.2^{+5.9}_{-8.4} \,{\rm MeV} \, ,
\end{eqnarray}
which is compatible with the mass and quantum numbers of the $Y(4500)$
recently detected by BESIII~\cite{BESIII:2022joj}.

For the $Y(4230)$, we notice that the $Y(4620)$ has been produced
in $e^{-} e^{+}$~\cite{Jia:2019gfe,Belle:2020wtd} and thus
we expect its quantum numbers to be $J^{PC} = 1^{--}$,
where we find a pole with mass
\begin{eqnarray}
  M(D_{s1} \bar{D}_s^*, 1^{--}) &=& {( 4622^{+22}_{-43} )}^V\,{\rm MeV} \, ,
\end{eqnarray}
with the superscript $V$ indicating a virtual state solution.
This virtual state has the correct quantum numbers to be a plausible
interpretation of the $Y(4620)$.
The problem is that virtual states are very difficult to observe unless
they are close to threshold, and in this case the center-of-mass
energy of this pole is $25^{+43}_{-22}\,{\rm MeV}$ below
the $D_s^* \bar{D}_{s1}$ threshold.
The uncertainties in the exact location of this virtual state are large though
and it might happen that it is closer to threshold than
our first approximation in the $Y$-counting suggests.
There are potentially important effects that we have not considered yet,
including the role of the $D_{0c}$ coupling or the coupled-channel
dynamics between the $D_s^* \bar{D}_{s1}$ and $D_s^* \bar{D}_{s2}^*$
channels, which are about $34\,{\rm MeV}$ away from each other.
As we will see, these effects will change the pole
into a virtual or bound state close to threshold.

Besides the $1^{--}$ $D_{s}^* \bar{D}_{s1}$ system, it is also interesting
to consider its $2^{--}$ counterpart, as their potential is the same
in the present counting, leading to the same mass prediction
in the same Riemann sheet:
\begin{eqnarray}
  M(D_{s1} \bar{D}_s^*, 2^{--}) &=& {( 4622^{+22}_{-43} )}^V\,{\rm MeV} \, .
\end{eqnarray}
However the most attractive configuration is the $3^{--}$ state, where
\begin{eqnarray}
  V_C^{\rm LO(Y)}(D_{s2}\bar{D}_s^*, 3^{--}) &=& - \frac{D_{0b}}{2} \, .
\end{eqnarray}
In Table \ref{tab:Y-counting} we predict the $3^{--}$ $D_{s2} \bar{D}_s^*$
system to be bound and have a mass of
\begin{eqnarray}
  M(D_{s2} \bar{D}_s^*, 3^{--} ) &=& 
  4640^{+14}_{-21}\,{\rm MeV} \, .
\end{eqnarray}
The prediction of three states ($1^{--}$, $2^{--}$ and $3^{--}$) with a mass
in the vicinity of $(4620-4640)\,{\rm MeV}$ will not change after
the inclusion of the second non-diagonal contact-range interaction ($D_{0c}$)
and coupled-channel effects, as we will see in the following lines.

\subsection{What about the $Y(4360)$?}

Another charmonium-like state which has been suspected to be molecular is
the $Y(4360)$, with a mass of $4382.0\,{\rm MeV}$ according to its latest
observation by BESIII~\cite{Ablikim:2020cyd}.
However it is difficult to predict in the $Y$-counting, where the reason is
that its ${\rm LO}$ potential reads
\begin{eqnarray}
  V_C^{\rm LO(Y)}(D^*\bar{D}_1, 1^{--}) &=& - \frac{D_{0b}}{6} \, ,
\end{eqnarray}
which is too weak to form a deep enough bound state.
There exists attraction though, with a virtual state being expected 
merely $2\,{\rm MeV}$ below threshold,
check Table~\ref{tab:Y-counting}.

This might change if we promote a second coupling to ${\rm LO}$.
In view of the phenomenological model of Ref.~\cite{Peng:2021hkr}
the $D_{0c}$ coupling seems a good choice,
with its promotion entailing a second power counting:
\begin{eqnarray}
  D_{0b} \, , \, D_{0c} \sim Q^{-1} \quad
  \mbox{and} \quad C_{0a} \, \dots \, D_{1c} \sim Q^0 \, ,
\end{eqnarray}
which we will call the $Y^*$-counting.
In this counting the ${\rm LO}$ potential in the $Y(4360)$ channel reads
\begin{eqnarray}
  V_C^{\rm LO(Y^*)}(D^*\bar{D}_1, 1^{--}) &=&
  - \frac{D_{0b}}{6} - \frac{5}{4}\,D_{0c}\, ,
\end{eqnarray}
where the new coupling $D_{0c}$ can be determined
from the mass of the $Y(4360)$.
Repeating the same steps as with the $Y(4230)$, we find
\begin{eqnarray}
  D_{0c} = 1.07\,(2.32-0.66)\,{\rm fm}^2 \, ,
\end{eqnarray}
for $\Lambda = 0.75\,{\rm GeV}$, where the values in parentheses
correspond to the $(0.5-1.0)\,{\rm GeV}$ range.

The ${\rm LO}$ potential for the other meson-antimeson configurations
and the predictions for this counting are shown
in Table \ref{tab:Yst-counting}.
It is interesting to notice the $J^{PC} = 1^{--}$ $D_{s}^* D_{s1}^*$ channel,
where the potential reads
\begin{eqnarray}
  V_C^{\rm LO(Y^*)}(D_{s}^* \bar{D}_{s1}, 1^{--}) &=&
  - \frac{D_{0b}}{12} - \frac{5}{8}\,D_{0c}\, ,
\end{eqnarray}
that is, half the strength of the potential in the $Y(4360)$ channel.
This time we predict a bound state at 
\begin{eqnarray}
  M(D_{s}^*  \bar{D}_{s1}, 1^{--}) = 4643.4^{+3.3}_{-6.0}\,{\rm MeV} \, ,
\end{eqnarray}
which is close to the experimental location of the $Y(4620)$.
Besides, there is a prediction of a $J^{PC} = 2^{--}$ $D_{s}^* D_{s1}$
virtual state close to threshold, with a mass
\begin{eqnarray}
  M(D_{s}^* \bar{D}_{s1}, 2^{--}) = {(4638.3^{+8.8}_{-22.3})}^V\,{\rm MeV} \, ,
\end{eqnarray}
and, as happened in the $Y$-counting, the most attractive hidden-strange
molecule is, as usual, the $3^{--}$ $D_{s}^* D_{s2}^*$
\begin{eqnarray}
  M(D_{s}^* \bar{D}_{s2}^*, 3^{--}) = 
  4624^{+19}_{-29}\,{\rm MeV} \, .
\end{eqnarray}
It is also worth noticing the existence of a $2^{-+}$ $D_{s}^* D_{s2}^*$
virtual bound state with mass
\begin{eqnarray}
  M(D_{s}^* D_{s2}^*, 2^{-+}) = {(4635^{+40}_{-82})}^V\,{\rm MeV} \, ,
\end{eqnarray}
which one might be tempted to identify with the $X(4630)$ state
found by the LHCb~\cite{Aaij:2021ivw}, except that it is unlikely
to be seen owing to its location in the complex plane.
Being so deep as to be below the $D_s^* \bar{D}_{s1}$ threshold,
this result only underscores that it might be advisable to include
coupled channel effects, which we will do in the following lines.

\begin{table*}[!tth]
  \begin{tabular}{|ccccc|ccccc|}
\hline\hline
  Molecule  & $J^{PC}$ & $V^{\rm LO(Y^*)}$ & B & M &
  Molecule  & $J^{PC}$ & $V^{\rm LO(Y^*)}$ & B & M \\
  \hline
  $D \bar{D}_1$  & $1^{-+}$ & $+\frac{2}{3}\,D_{0b}$ & - & - &
  $D_s \bar{D}_{s1}$  & $1^{-+}$ & $+\frac{1}{3}\,D_{0b}$ & - & - \\
  $D \bar{D}_1$ &  $1^{--}$ & $-\frac{2}{3}\,D_{0b}$ & 
  Input ($70.7$) & Input ($4218.6$) & 
  $D_s \bar{D}_{s1}$  & $1^{--}$ & $-\frac{1}{3}\,D_{0b}$ &
  {$9.3^{+8.7}_{-6.0}$} & {$4494.2_{-8.7}^{+6.0}$} \\
  \hline
  $D \bar{D}_2^*$  & $2^{-+}$ & $+D_{0c}$ & - & - &
  $D_s \bar{D}_{s2}^*$  & $2^{-+}$ & $+\frac{1}{2}\,D_{0c}$ & - & - \\
  $D \bar{D}_2^*$  & $2^{--}$ & $-D_{0c}$ &
  {$5.1^{+6.0}_{-3.8}$} & {$4323.2_{-6.0}^{+3.8}$} &
  $D_s \bar{D}_{s2}^*$  & $2^{--}$ & $-\frac{1}{2}\,D_{0c}$ &
  $3.6_{-3.5}^{+12.7}$ & ${4533.9^{+3.5}_{-12.7}}$ \\
  \hline
  $D^* \bar{D}_1$ &   $0^{-+}$ & $- \frac{1}{3}\,D_{0b} + \frac{5}{2}\,D_{0c}$
  & - & - &
  $D_s^* \bar{D}_{s1}$  & $0^{-+}$ &
  $- \frac{1}{6}\,D_{0b} + \frac{5}{4}\,D_{0c}$ & - & - \\
  $D^* \bar{D}_1$ &   $0^{--}$ & $+ \frac{1}{3}\,D_{0b} - \frac{5}{2}\,D_{0c}$ &
  ${26.0^{+13.0}_{-8.7}}$ & ${4404.6^{+8.7}_{-13.0}}$ & 
  $D_s^* \bar{D}_{s1}$  & $0^{--}$ &
  $+ \frac{1}{6}\,D_{0b} - \frac{5}{4}\,D_{0c}$ &
  ${0.2^{+2.7}_{-0.2}}$ & ${4647.1_{-2.7}^{+0.2}}$\\
  $D^* \bar{D}_1$ &   $1^{-+}$ &
  $+ \frac{1}{6}\,D_{0b} + \frac{5}{4}\,D_{0c}$ & - & - &
  $D_s^* \bar{D}_{s1}$   & $1^{-+}$ &
  $+ \frac{1}{12}\,D_{0b} + \frac{5}{8}\,D_{0c}$ & - & - \\
  $D^* \bar{D}_1$ &   $1^{--}$ & $- \frac{1}{6}\,D_{0b} - \frac{5}{4}\,D_{0c}$ &
  Input ($48.7$) &  Input($4382.0$) &
  $D_s^* \bar{D}_{s1}$  &  $1^{--}$
  & $- \frac{1}{12}\,D_{0b} - \frac{5}{8}\,D_{0c}$ &
  $3.9^{+6.0}_{-3.3}$ & $4643.4^{+3.3 }_{-6.0}$
  \\
  $D^* \bar{D}_1$ &   $2^{-+}$ &
  $+ \frac{1}{6}\,D_{0b} + \frac{1}{4}\,D_{0c}$ & - & - &
  $D_s^* \bar{D}_{s1}$  &  $2^{-+}$
  & $+ \frac{1}{12}\,D_{0b} + \frac{1}{8}\,D_{0c}$ & - & - \\
  $D^* \bar{D}_{1}$  &  $2^{--}$ &
  $- \frac{1}{6}\,D_{0b} - \frac{1}{4}\,D_{0c}$ &
  ${0.70^{+5.4}_{-0.7}}$ & ${4430.0_{-5.4}^{+0.7}}$ &
  $D_s^* \bar{D}_{s1}$  &  $2^{--}$ &
  $- \frac{1}{12}\,D_{0b} - \frac{1}{8}\,D_{0c}$ &
  ${(9.0_{-8.8}^{+22.3})}^V$ & ${\left(4638.3^{+8.8}_{-22.3} \right)^V}$ \\
  \hline
  $D^* \bar{D}_2^*$ &   $1^{-+}$ & $+ \frac{1}{6}\,D_{0b} - \frac{3}{4}\,D_{0c}$ &
  ${{(42_{-28}^{+44})}^V}$ &${\left(4428^{+28}_{-44}  \right)^V}$ &
  $D_s^* \bar{D}_{s2}^*$   & $1^{-+}$ &
  $+ \frac{1}{12}\,D_{0b} - \frac{3}{8}\,D_{0c}$ &
  ${{(84_{-52}^{+79})}^V}$ & ${\left(4597^{+52}_{-79} \right)^V}$ \\
  $D^* \bar{D}_2^*$ &  $1^{--}$ &
  $- \frac{1}{6}\,D_{0b} + \frac{3}{4}\,D_{0c}$ & - & - &
  $D_s^* \bar{D}_{s2}^*$  & $1^{--}$ &
  $- \frac{1}{12}\,D_{0b} + \frac{3}{8}\,D_{0c}$ & - & - \\
  $D^* \bar{D}_2^*$ &   $2^{-+}$ &
  $ - \frac{1}{2}\,D_{0b} + \frac{5}{4}\,D_{0c}$ &
  ${(14_{-14}^{+44})}^V$ & ${\left(4456^{+14}_{-44} \right)^V}$ &
  $D_s^* \bar{D}_{s2}^*$  & $2^{-+}$ &
  $ - \frac{1}{4}\,D_{0b} + \frac{5}{8}\,D_{0c}$ &
  ${(46_{-40}^{+82})}^V$ & ${\left(4635^{+40}_{-82} \right)^V}$ \\
  $D^* \bar{D}_2^*$ &   $2^{--}$ &
  $+ \frac{1}{2}\,D_{0b} - \frac{5}{4}\,D_{0c}$ & & &
  $D_s^* \bar{D}_{s2}^*$  & $2^{-+}$ &
  $ - \frac{1}{4}\,D_{0b} - \frac{5}{8}\,D_{0c}$ & - & - \\
  $D^* \bar{D}_2^*$ &   $3^{-+}$ & $+ D_{0b} + \frac{1}{2}\,D_{0c}$ & - & - &
  $D_s^* \bar{D}_{s2}^*$   & $3^{-+}$ &
  $+ \frac{1}{2}\,D_{0b} + \frac{1}{4}\,D_{0c}$ & - & - \\
  $D^* \bar{D}_2^*$ &  $3^{--}$ & $- D_{0b} - \frac{1}{2}\,D_{0c}$ &
  ${200_{-96}^{+172}}$ & ${4270^{+96}_{-172}}$ & 
  $D_s^* \bar{D}_{s2}^*$  & $3^{--}$ &
  $- \frac{1}{2}\,D_{0b} - \frac{1}{4}\,D_{0c}$ &
  ${58^{+29}_{-19}}$  & ${4624^{+19}_{-29}}$ \\ 
  \hline\hline 
\end{tabular}
  \caption{
    Contact-range potential and bound state spectrum in the $Y^*$-counting,
    the second counting we are considering here.
    The assumptions behind this counting are that the $Y(4230)$ and $Y(4360)$
    are a $J^{PC} = 1^{--}$ $D \bar{D}_1$ and $D^* \bar{D}_1$ bound states
    generated by the strength of the $D_{b0}$ and $D_{c0}$ couplings
    (while all the other couplings are expected to be small
    in comparison).
    The central predictions correspond to the cutoff
    $\Lambda = 0.75\,{\rm GeV}$, while the error is the maximum
    of either the cutoff dependence in the $(0.5-1.0)\,{\rm GeV}$ window or
    the intrinsic EFT truncation error, see Eq.~(\ref{eq:B-err}) and
    the related discussion for details.
    The superscript $V$ indicates a virtual state solution
    (instead of a bound state one).
    The binding energies and masses are in units of ${\rm MeV}$.
}
\label{tab:Yst-counting}
\end{table*}

\subsection{Coupled channel effects}
\label{subsec:CC-effects}

The mass difference of the ground and excited states of the P-wave charmed
mesons is small in comparison with their S-wave counterparts.
Indeed, we have that
\begin{eqnarray}
  m(D_{2}^*) - m(D_1) = 39 \,\, , \,\,
  m(D_{s2}^*) - m(D_{s1}) = 34\,{\rm MeV} \, , \nonumber \\
\end{eqnarray}
respectively, which happens to be comparable with the binding energies of
the states we have been predicting in Tables \ref{tab:Y-counting} and
\ref{tab:Yst-counting}.
In terms of a momentum scale for the ${D}^* \bar{D}_{2}^*$ and
${D}_s^* \bar{D}_{s2}^*$ systems, the previous mass differences
would correspond with a coupled-channel momentum of
$\Lambda_{\rm CC} = 288$ and $275\,{\rm MeV}$,
which belongs to the light scales.
This suggests that the inclusion of coupled channel effects
would be a possible improvement to our EFT description.

The molecular configurations affected by coupled channel effects are
the $J = 1, 2$ ${D}^* \bar{D}_{1}$-${D}^* \bar{D}_{2}^*$ and
${D}_s^* \bar{D}_{s1}$-${D}_s^* \bar{D}_{s2}^*$ systems.
If we limit ourselves to the scope of the two power countings proposed
in this work, the coupled channel potentials that we will obtain
in these configurations for the non-strange sector are
\begin{eqnarray}
  && V_C^{{\rm LO} (Y/Y^*)}(I=0, J^{PC} = 1^{-\pm}) = \nonumber \\
  && \quad \pm\,
  \begin{pmatrix}
    \frac{1}{6} D_{0b} + \frac{5}{4} D_{0c} &
    \frac{\sqrt{5}}{6} D_{0b} - \frac{3 \sqrt{5}}{4} D_{0c} \\
    \frac{\sqrt{5}}{6} D_{0b} - \frac{3\sqrt{5}}{4} D_{0c} &
    \frac{1}{6} D_{0b} - \frac{3}{4} D_{0c}
  \end{pmatrix} \, , \label{eq:VY1-CC} \\
  \nonumber \\
  && V_C^{{\rm LO} (Y/Y^*)}(I=0, J^{PC} = 2^{-\pm}) = \nonumber \\
  && \quad \pm\,
  \begin{pmatrix}
    \frac{1}{6} D_{0b} + \frac{1}{4} D_{0c} &
    -\frac{1}{2} D_{0b} -  \frac{3}{4} D_{0c} \\
    -\frac{1}{2} D_{0b} - \frac{3}{4} D_{0c}  &
    -\frac{1}{2} D_{0b} + \frac{5}{4} D_{0c}
  \end{pmatrix} \, , \label{eq:VY2-CC}
\end{eqnarray}
where the sign depends on the C-parity configuration,
while for the hidden-strange sector the potentials
will be half the previous ones
\begin{eqnarray}
  && V_C^{{\rm LO} (Y/Y^*)}(s\bar{s}, J^{PC} = 1^{-\pm}) = \nonumber \\
  && \qquad
  \frac{1}{2}\, V_C^{{\rm LO} (Y/Y^*)}(I=0, J^{PC} = 1^{-\pm})
  \, , \label{eq:VY1-SS} \\
  && V_C^{{\rm LO} (Y/Y^*)}(s\bar{s}, J^{PC} = 2^{-\pm}) = \nonumber \\
  && \qquad
  \frac{1}{2}\, V_C^{{\rm LO} (Y/Y^*)}(I=0, J^{PC} = 2^{-\pm})
  \, . \label{eq:VY2-SS} 
\end{eqnarray}
The most important effect of these potentials is to favor the formation of
bound states in the $J=1,2$ ${D}^* \bar{D}_{1}$ and ${D}_s^* \bar{D}_{s1}$
systems, while disfavoring their higher mass counterparts
$J=1,2$ ${D}^* \bar{D}_{2}^*$ and ${D}_s^* \bar{D}_{s2}^*$
(which actually disappear).
In Table~\ref{tab:CC-counting} we summarize the new predictions
for the ${D}^* \bar{D}_{1}$ and ${D}_s^* \bar{D}_{s1}$ systems
when coupled channels are included
in the $Y$- and $Y^*$-countings.

The inclusion of these coupled channels does not modify
the determination of the $D_{0b}$ coupling in the $Y$-counting,
only its predictions for the affected molecular configurations.
The most significant changes happen in the $1^{--}$ $D^* \bar{D}_1$ channel,
where the new prediction is
\begin{eqnarray}
  M(D^*\bar{D}_1, 1^{--}) = 
  4401.0^{+9.9}_{-14.8}\,{\rm MeV} \, ,
\end{eqnarray}
and the appearance of poles with positive C-parity, which might provide a
plausible molecular interpretation for the $X(4630)$
\begin{eqnarray}
  M(D_s^*\bar{D}_{s1}, 1^{-+}) &=&
  {( 4644^{+3}_{-28} )}^{V'} \, , \\
  M(D_s^*\bar{D}_{s1}, 2^{-+}) &=&
  4647^{+19 - i 17}_{-29} \,{\rm MeV} \, .
\end{eqnarray}
The first configuration is a virtual state with respect to
the $D_s^* D_{s1}$ and $D_s^* D_{s2}^*$ thresholds, that is,
the (II,II) Riemann sheet (denoted with the superscript ${V'}$),
from which we do not expect it to be observable in experiments.
The second configuration generates a bound state basically at threshold,
which might correspond with the $X(4630)$, though depending on the cutoff
it might become more or less bound, where in this later case it can
move to the (II,I) Riemann sheet (i.e. becoming a resonance).

For the $Y^*$-counting, the situation is slightly different.
As a consequence of the new channel
we have to recalibrate the $D_{0c}$ coupling, resulting in
\begin{eqnarray}
  D_{0c} = 1.02\,\left(2.24-0.62 \right) \,{\rm fm}^2 \, ,
\end{eqnarray}
that is, a slightly weaker coupling (now the higher mass channel provides
additional attraction to this system).
In this counting the $J^{PC} = 1^{-+}$ $D_s^*\bar{D}_{s1}$ configuration is more
attractive than before, with the pole moving to the (I,II) Riemann
sheet and located at
\begin{eqnarray}
  M(D_s^*\bar{D}_{s1}, 1^{-+})
  &=& {(4594^{+59 - i 1}_{-89})}^{V^*}\,{\rm MeV} \, , 
\end{eqnarray}
where the ${V^*}$ superscript indicates the previous combination of
sheets in the complex plane.
Being in the (I,II) sheet, the $1^{-+}$ state is still too
deep to be observable and hence the $Y^*$-counting with
coupled channels still disfavors interpreting it
as the $X(4630)$.
In contrast, the $2^{-+}$ configuration still binds and
happens to be located basically at threshold, again:
\begin{eqnarray}
  M(D_s^*\bar{D}_{s1}, 2^{-+}) = 
  4647.3_{-4.3 }^{+0.0}\,{\rm MeV} \, ,
\end{eqnarray}
where the errors are asymmetric as a consequence that for softer (harder)
cutoffs it increases binding (becomes a virtual state below threshold).
The bound state solution is not far away from the experimental mass of
the $X(4630)$ (and compatible once we take into account the experimental
uncertainty, i.e. $4626^{+18}_{-110}\,{\rm MeV}$), with its preferred
spin-parity $1^-$ or $2^-$.
Thus the $X(4630)$ can potentially correspond with
our predicted $2^{-+}$ $D_s^*\bar{D}_{s1}$ state.

\begin{table*}[!tth]
  \begin{tabular}{|ccccc|ccccc|}
\hline\hline
  Molecule & $J^{PC}$ & $V^{\rm LO(Y)}$ & B & M &
  Molecule & $J^{PC}$ & $V^{\rm LO(Y)}$ & B & M \\
  \hline
  $D^* \bar{D}_1$-$D^* \bar{D}_2^*$  & $1^{-+}$ & Eq.~(\ref{eq:VY1-CC}) &
  - & - &
  $D_s^* \bar{D}_{s1}$-$D_s^* \bar{D}_{s2}^*$  & $1^{-+}$ & Eq.~(\ref{eq:VY1-SS})
  & - & - \\
  $D^* \bar{D}_1$-$D^* \bar{D}_2^*$  & $1^{--}$ & Eq.~(\ref{eq:VY1-CC}) &
  $29.6^{+14.8}_{-9.9}$ & $4401.0^{+9.9}_{-14.8}$ & 
  $D_s^* \bar{D}_{s1}$-$D_s^* \bar{D}_{s2}^*$  & $1^{--}$ & Eq.~(\ref{eq:VY1-SS})
  & ${(1.0_{-1.0}^{+14.9})}^V$ & ${(4646.3^{+1.0}_{-14.0})}^V $ \\
  $D^* \bar{D}_1$-$D^* \bar{D}_2^*$  & $2^{-+}$ & Eq.~({\ref{eq:VY2-CC}}) &
  $70^{+35}_{-23}$ & $4361^{+23}_{-35}$ & 
  $D_s^* \bar{D}_{s1}$-$D_s^* \bar{D}_{s2}^*$  & $2^{-+}$ & Eq.~({\ref{eq:VY2-SS}}) &
  ${0.0}^{+4.7}_{-8.1 + 16.7 i}$ & ${4647.3^{+8.1}_{-4.7}}$ \\  
  $D^* \bar{D}_1$-$D^* \bar{D}_2^*$  & $2^{--}$ & Eq.~({\ref{eq:VY2-CC}}) &
  $18.3^{+9.2}_{-6.5}$ & $4412.4_{-9.2}^{+6.5}$ & 
  $D_s^* \bar{D}_{s1}$-$D_s^* \bar{D}_{s2}^*$  & $2^{--}$ & Eq.~({\ref{eq:VY2-SS}}) &
  ${(1.2_{-1.2}^{+12.3})^V}$ & ${(4646.0^{+1.2}_{-12.3})}^V$ \\
  \hline \hline
  Molecule & $J^{PC}$ & $V^{\rm LO(Y^*)}$ & B & M &
  Molecule & $J^{PC}$ & $V^{\rm LO(Y^*)}$ & B & M \\
  \hline
  $D^* \bar{D}_1$-$D^* \bar{D}_2^*$  & $1^{-+}$ & Eq.~(\ref{eq:VY1-CC}) &
    -  & - &
    $D_s^* \bar{D}_{s1}$-$D_s^* \bar{D}_{s2}^*$  & $1^{-+}$ & Eq.~(\ref{eq:VY1-SS})
    & - & - \\
  $D^* \bar{D}_1$-$D^* \bar{D}_2^*$  & $1^{--}$ & Eq.~(\ref{eq:VY1-CC}) &
  Input($48.7$) & Input($4382.0$) &
  $D_s^* \bar{D}_{s1}$-$D_s^* \bar{D}_{s2}^*$  & $1^{--}$ & Eq.~(\ref{eq:VY1-SS})
  & $3.7^{+6.1}_{-3.2}$ & $4643.7_{-6.1}^{+3.2}$ \\
  $D^* \bar{D}_1$-$D^* \bar{D}_2^*$  & $2^{-+}$ & Eq.~(\ref{eq:VY2-CC}) &
  $57^{+29}_{-19}$ & $4374^{+19}_{-29}$ &
  $D_s^* \bar{D}_{s1}$-$D_s^* \bar{D}_{s2}^*$  & $2^{-+}$ & Eq.~(\ref{eq:VY2-SS}) &
  $0.0^{+4.3}_{-0.0}$ & $4647.3^{+0.0}_{-4.3}$ \\
  $D^* \bar{D}_1$-$D^* \bar{D}_2^*$  & $2^{--}$ & Eq.~(\ref{eq:VY2-CC}) &
  $93_{-34}^{+55}$ & $4338^{+34}_{-55}$ & 
  $D_s^* \bar{D}_{s1}$-$D_s^* \bar{D}_{s2}^*$  & $2^{--}$ & Eq.~(\ref{eq:VY2-SS}) &
  $11.7^{+6.2}_{-5.1}$ & $4635.6_{-6.2}^{+5.1}$ \\
  \hline
  \end{tabular}
  \caption{Predictions for the $D^* \bar{D}_{1}$ and $D_s^* \bar{D}_{s1}$
    molecules in the $Y$- and $Y^*$-countings once the coupled channel
    dynamics with the $D^* \bar{D}_{2}$ and $D_s^* \bar{D}_{s2}^*$
    thresholds is included.
    Binding energies and masses are in units of ${\rm MeV}$.
    The superscript $V$ indicates a state in the (II,I) Riemann sheet
    (with respect to the $D^* \bar{D}_{1}$-$D^* \bar{D}_{2}$ and
    $D_s^* \bar{D}_{s1}$-$D_s^* \bar{D}_{s2}^*$ threshold,
    depending on whether we are in the non-strange or strange sectors),
    while no superscript refers to the (I,I) sheet.
    For the $1^{-+}$ configurations, there is attraction but the poles happen
    in the (I,II) and (II,II) Riemann sheets, far from physical scattering;
    we do not list the $1^{-+}$ molecules in this Table, but discuss
    them briefly in Sect.~\ref{subsec:CC-effects}.
  }
  \label{tab:CC-counting}
\end{table*}

\subsection{Limitations of the present EFT approach}

Both the $Y$- and $Y^*$-counting are simplifications of the complex operator
structure of the contact-range potential for this type of molecules.
Despite having five candidates, only two of them have clear molecular
interpretations, the $Y(4230)$ and $Y(4360)$; two of the hidden-strange
candidates seemingly are the SU(3)-flavor partners of the previous two
non-strange states, the $Y(4500)$ and $Y(4620)$ that might be partner
states of the $Y(4230)$ and $Y(4360)$, respectively.
Meanwhile, the hidden-strange $X(4630)$ state is more difficult to interpret
without coupled-channel dynamics.

The previous limitations, combined with the phenomenological observation that
the E1 term (and to a lesser extent the M2 term) of the S-wave potential
might be relatively strong, lead to the choices behind the $Y$- and
$Y^*$-countings.
These two countings are seemingly able to explain the five previous states,
specially when coupled channel dynamics are included.
Yet, they are incomplete.
Luckily it is not difficult to analyze in which molecular configurations
the previous two countings are more likely to fail.

The crucial assumption of the $Y$- and $Y^*$-counting is the following
ordering of the size of the couplings
\begin{eqnarray}
  | D_{0b} | > |D_{0c} | > |C_{0a}| , |C_{0b}| \, ,
\end{eqnarray}
for which there is circumstantial evidence from saturating the value of
these couplings with scalar and vector meson exchange and
then estimating the magnitude of the later 
from vector meson dominance or the condition of
reproducing the $Y(4230)$ and $Y(4360)$~\cite{Peng:2021hkr}.
The obvious problem is that the previous ordering only works when the numerical
factors in front of all the couplings is of $\mathcal{O}(1)$.
Thus, the $Y$- and $Y^*$-countings are expected to work well
for the $D \bar{D}_1$, $D \bar{D}_2^*$ and
$J=3$ $D^* \bar{D}_2^*$ systems,
while their applicability in other configurations will be more limited.

This will generate systematic deviations from our predictions, which can be 
estimated up to a certain extent.
For instance, the $C_{Ia}$ couplings are more likely to be attractive
than repulsive and are also likely to be attractive in both $I=0,1$
configurations (with $I=0$ being the most attractive one):
\begin{eqnarray}
  C_{Ia} < 0 \quad \mbox{and} \quad  |C_{0a}| > |C_{1a}| \, .
\end{eqnarray}
The reasons behind this hypothesis are on the one hand the saturation of
the couplings from scalar and vector meson exchange and on the other
the fact that similar relations can be deduced from the molecular
candidates in the $D^{(*)}\bar{D}^{(*)}$ system, e.g. the $X(3872)$
and $Z_c(3900/4020)$, where caveats might apply
as these are different systems and their
couplings are not the same.
If the previous relation happens to be the case, this indicates that we are
underestimating the magnitude of the attraction in configurations
where the coefficient in front of $D_{0b}$ and $D_{0c}$ is small
and also in the hidden-strange configurations.
Thus, this argument points towards more binding for the hidden-strange
molecules and for most of the $D^* \bar{D}_1$ and $D^* \bar{D}_1^*$
configurations.

The next deviation regards the spin-spin coupling $C_{0b}$,
for which the sign is not known for sure.
There exist two hypotheses in the literature:
\begin{eqnarray}
  C_{0b} > 0 \quad \mbox{and} \quad C_{0b} < 0 \, ,
\end{eqnarray}
which we will call scenarios A and B (in analogy with how these possibilities
have been previously named in the pentaquark case~\cite{Liu:2019tjn}).
Scenario A generates a spectrum in which the molecules with higher light-spin
content are heavier, while scenario B predicts the opposite pattern. 
Scenario A is the standard one for compact hadrons, but it might also happen
in isoscalar two-hadron states if the contact-range couplings are saturated
by the direct interaction of the light-quarks within the hadrons,
as proposed in Ref.~\cite{Chen:2021cfl}.
Instead, if the contact-range couplings are saturated
by light-meson exchange~\cite{Peng:2020xrf} (plus the assumption that
the Dirac-delta terms appearing in the spin-spin vector-meson
exchange terms can be neglected at the scales relevant
for the formation of bound states, see e.g. the brief discussion
after Eqs.~(\ref{eq:Va}-\ref{eq:Wc}) or the ones
in Refs.~\cite{Liu:2019zvb,Yalikun:2021bfm}
for the pentaquark case),
then the expected outcome will be scenario B.
Which of these two scenarios is correct matters for the molecular
interpretation of the $X(4630)$, in particular whether it is a
$1^{-+}$ or $2^{-+}$ state.
The reason is that in these two channels the potential reads
\begin{eqnarray}
  V_C(D^*\bar{D}_{1}, 1^{-+}) = C_{a} - \frac{5}{4} C_{b} + \frac{1}{6} D_{b} + \frac{5}{4} D_{c} \, , \\
  V_C(D^*\bar{D}_{1}, 2^{-+}) = C_{a} + \frac{5}{4} C_{b} + \frac{1}{6} D_{b} + \frac{1}{4} D_{c} \, , 
\end{eqnarray}
where for simplicity we have not specified the isospin or flavor representation.
If scenario $A$ is correct, the $1^{-+}$ configuration (which is already
attractive, only not enough as to generate a pole near threshold)
will be more attractive than expected, while the contrary
will happen for the $2^{-+}$ configuration.
This might potentially change our conclusions and indicate that the most
promising molecular interpretation of the $X(4630)$ is a
$1^{-+}$ $D_{s}^* \bar{D}_{s1}$ bound state, which will be
in agreement with Ref.~\cite{Yang:2021sue}.
Conversely, if scenario $B$ is correct, this will further cement
the conclusion that the $X(4630)$ is better reproduced
as a $2^{-+}$ $D_{s}^* D_{s1}$ bound state.
  In the next lines we will explicitly consider two phenomenological models
  as a way to further understand the molecular spectrum and its relation
  with the factors we have discussed here.

\subsection{Comparison with phenomenology}
\label{subsec:comparison}

Actually, there is a possible way to address the unknown factors that
we have discussed in the previous subsection: to calculate the molecular
spectrum in a phenomenological model and compare
these predictions with the EFT ones.

Even though phenomenology is intrinsically model-dependent, it nonetheless
contains all the interaction structures we have discussed ($V_a$, $V_c$,
$W_b$, $W_c$ or their contact-range EFT counterparts $C_a$, $C_b$,
$D_b$, $D_c$), including those that we are currently unable to
determine owing to the absence of suitable molecular
candidates (i.e. $C_a$ and $C_b$).
In this regard, if EFT and phenomenological predictions happen to converge,
that might indicate that we are on the right track for determining
the spectrum of S- and P-wave charmed meson molecules.

For this comparison we have chosen the following two phenomenological models:
\begin{itemize}
\item[(i)] The one boson exchange (OBE)
  model~\cite{Machleidt:1987hj,Machleidt:1989tm},
  in which the potential among hadrons
  is the sum of the potential generated by the exchange of a few
  light-mesons, in our case the scalar and vector meson
  exchange potential of Eqs.~(\ref{eq:Va}-\ref{eq:Wc}),
  but modified by the inclusion of finite-size effects
  (i.e. form factors). 
\item[(ii)] A saturation model in which the finite-range potential
  generated by the exchange of a light meson is simplified to a
  contact-range potential with a coupling strength proportional
  to that of the original finite-range potential.
  By means of a renormalization group (RG) equation, the previous coupling
  strength is in turn modified as to take into account the different
  ranges of the scalar and vector mesons~\cite{Peng:2021hkr},
  which is why we will refer to this model as
  the RG-saturation model.
\end{itemize}
We describe the technical details of these models
in Appendix~\ref{app:phenomenology}, while here
we concentrate on comparing their results to those of EFT.
Yet, we will briefly comment on a few commonalities of the two models.
First, in the OBE model there is an unknown form factor and cutoff,
while in the RG-saturation model there is an unknown proportionality
constant for the saturated coupling.
In both cases we determine these unknown factors by reproducing
the $X(3872)$ as a $J^{PC} = 1^{++}$, $I=0$ $D^*\bar{D}$
molecule (that is, we are assuming a molecular $X(3872)$).
Second, while most of the couplings of the scalar and vector mesons are
relatively well known, this is not the case for the $E1$ and $M2$
components of the vector meson potential, i.e. the $f_V'$ and $h_V'$
couplings of Eqs.~(\ref{eq:Wb}) and (\ref{eq:Wc}).
Here we fix $f_V'$ and $h_V'$ from the condition of reproducing
the masses of the $Y(4230)$ and $Y(4360)$.

We also provide error estimations for the OBE and RG-saturation models.
The status and parameters of the scalar meson are much more uncertain than
those of the vector mesons. In particular, the scalar meson mass is not
well known, where the RPP cites the $m_S = (400-550)\,{\rm MeV}$
range~\cite{Zyla:2020zbs}.
This uncertainty, which by itself implies a variability of the overall strength
of the potential of the order of $(30-40)\%$, can be propagated to
the predictions of both models.
For the RG-saturation model a second source of uncertainty is that it uses
the contact-range approximation, which is a good approximation only if
the predicted bound state is sufficiently large in size.
Otherwise the wave function of the bound state will be able to probe
the short-range details of the interaction binding them.
For this we will consider the same type of $\gamma / m_{V}$ relative error
that we already included in the EFT calculations, i.e.
$[(1+\frac{\gamma_{\rm max}}{m_V})^{\pm 1} - 1]$ with $\gamma_{\rm max}$
the largest of the relevant binding momenta (either of the input states,
i.e. the $Y(4230)$, or the predicted states),
check Eqs.~(\ref{eq:B-err}) and (\ref{eq:B-intrinsic})
and the related discussion for details.

The comparison among these two models and the two power countings
we have discussed is shown in Tables \ref{tab:predictions-nonstrange} and
\ref{tab:predictions-strange} for the non-strange and hidden strange sectors.
It is important to notice that all calculations include
the $D_1 \bar{D}^*$-$D_2^* \bar{D}^*$ and
$D_{s1} \bar{D}_s^*$-$D_{s2}^* \bar{D}_s^*$ coupled channel effects when relevant.
From inspecting Tables \ref{tab:predictions-nonstrange} and
\ref{tab:predictions-strange}, it can be appreciated that
the two phenomenological models tend to predict a very
similar set of molecular states.
There are exceptions though, the most notable one being the $J^{PC} = 1^{-+}$
$D_1 \bar{D}^*$ and $D_{s1} \bar{D}_s^*$ configurations, which are predicted
as unbound in the OBE model but as bound in the RG-saturation model.
This configuration is interesting as it might correspond to the $X(4630)$,
for which the spin is either $J=1$ or $2$, and in this regard
the RG-saturation model predicts a candidate state
for both spins.

For the rest of the states, the predictions of both phenomenological models
basically coincide with the ones of the $Y^*$-counting, while there are
disagreements with the $Y$-counting for the $D_2^* \bar{D}$ and
$D_{s2}^* \bar{D}_s$ configurations, for the simple reason that
the coupling providing binding for these molecules is missing,
and also the $J=0$ $D_1 \bar{D}^*$ and $D_{s1} \bar{D}_s^*$
molecules.
In this last case, the $Y$- and $Y^*$-countings predict binding
for the $J^{PC} = 0^{-+}$ and $0^{--}$, respectively.
However, as we explained in the previous subsection, the predictions
for these configurations are uncertain, particularly
in the $Y$-counting.
The comparison with the two phenomenological models presented here supports
the previous conclusion, except that for the $Y^*$-counting the predictions
basically align with phenomenology, and also with the recent prediction of
a $J^{PC} = 0^{--}$ $D_1 \bar{D}^*$ state in Ref.~\cite{Ji:2022blw}.
Yet, caution is advised in view of the discrepancies in the prediction of
its mass in the OBE and RG-saturation models and EFT, suggesting
large uncertainties.

\begin{table*}[!ttt]
\begin{tabular}{|cccccccc|}
\hline\hline
Molecule  & $J^{PC}$ & $M_{\rm OBE}$ & $M_{\rm sat}$ & $M_Y$ & $M_{Y^*}$ & Candidate & $M_{\rm cand}$ \\
  \hline
  $D \bar{D}_1$ & $1^{-+}$ & $-$ & $-$ & $-$ & $-$ & ? & ? \\
  $D \bar{D}_1$ & $1^{--}$ & Input & Input & Input & Input & $Y(4230)$ & $4218.6 \pm 4.5$~\cite{Ablikim:2020cyd} \\
  \hline
  $D \bar{D}_2^*$ & $2^{-+}$ & $-$ & $-$ & $-$ & $-$ & ? & ? \\
  $D \bar{D}_2^*$ & $2^{--}$ & $4259.1 \pm 0.0$ & $4288^{+14}_{-20}$ & - & {$4323.2_{-6.0}^{+3.8}$}  & ? & ? \\
  \hline
  $D^* \bar{D}_1$ & $0^{-+}$ & $-$ & $-$ & $4421.7^{+5.9}_{-8.7}$ & $-$ & ? & ? \\
  $D^* \bar{D}_1$ & $0^{--}$ &
  $4393.5^{+3.9}_{-5.2}$ & $4430.2^{+0.5}_{-2.6}$ & $-$ & ${4404.6_{-13.0}^{+8.7}}$ & ? & ? \\
  $D^* \bar{D}_1$ & $1^{-+}$ & $-$ & $4357^{+25}_{-39}$ & $-$ & $-$ & ? & ? \\
  $D^* \bar{D}_1$ & $1^{--}$ & Input & Input & $4401.0^{-9.9}_{+14.8}$ & Input & $Y(4360)$  & $4382.0 \pm 13.4$~\cite{Ablikim:2020cyd} \\
  $D^* \bar{D}_1$ & $2^{-+}$ & $4387^{+13}_{-10}$ & $4282^{+61}_{-101}$ & $4361^{+23}_{-35}$ & $4374^{+19}_{-29}$ & ? & ? \\
  $D^* \bar{D}_1$ & $2^{--}$ & $4358^{+12}_{-11}$ & $4380^{+17}_{-26}$ & $4412.4^{+6.5}_{-9.2}$ & $4338^{+34}_{-55}$ & ? & ? \\
  \hline
  $D^* \bar{D}_2^*$ & $1^{-+}$ & $-$ & $-$ & $-$ & $-$ & ? & ? \\
  $D^* \bar{D}_2^*$ & $1^{--}$ & $-$ & $-$ & $-$ & $-$ & ? & ? \\
  $D^* \bar{D}_2^*$ & $2^{-+}$ & $-$ & $-$ & $-$ & $-$ & ? & ? \\
  $D^* \bar{D}_2^*$ & $2^{--}$ & $-$ & $4453.9^{+5.3}_{-7.9}$ & $-$ & $-$ & ? & ? \\
  $D^* \bar{D}_2^*$ & $3^{-+}$ & $-$ & $-$ & $-$ & $-$ & ? & ? \\
  $D^* \bar{D}_2^*$ & $3^{--}$ & $4323^{+28}_{-24}$ & $4302^{+74}_{-133}$ & $4311^{+68}_{-121}$ & ${4270^{+96}_{-172}}$ & ? & ? \\
  \hline \hline
\end{tabular}
\caption{Predictions from two phenomenological models --- the OBE model and
  the RG-saturation model --- and comparison with the predictions in power
  countings $Y$ and $Y^*$.
  For the phenomenological models, we have calibrated the electric dipolar and
  magnetic quadrupolar couping constants of the vector mesons to the condition
  of reproducing the $Y(4230)$ and $Y(4360)$ states.
  The calculations for the $D^* \bar{D}_1$ and $D^* \bar{D}_2^*$ systems
  contain the coupled channel effects.
  For further details, we refer to the description of these models
  in the main text and in Appendix \ref{app:phenomenology}.
}
\label{tab:predictions-nonstrange}
\end{table*}

\begin{table*}[!ttt]
\begin{tabular}{|cccccccc|}
  \hline \hline
Molecule  & $J^{PC}$ & $M_{\rm OPE}$ &  $M_{\rm sat}$  & $M_Y$ & $M_{Y^*}$ & Candidate & $M_{\rm cand}$ \\
  \hline
  $D_s \bar{D}_{s1}$ & $1^{-+}$ & $-$ & $-$ & $-$ & $-$ & ? & ? \\
  $D_s \bar{D}_{s1}$ & $1^{--}$ & $4503.9^{+0.4(V)}_{-1.8}$ & ${(4503.4^{0.1(B)}_{-1.6})}^V$ & $4494.2^{+5.9}_{-8.4}$ & {$4494.2_{-8.7}^{+6.0}$} & $Y(4500)$ & $4484.7 \pm 27.5$~\cite{BESIII:2022joj} \\
  \hline
  $D_s \bar{D}_{s2}^*$ & $2^{-+}$ & $-$ & $-$ & $-$ & $-$ & ? & ? \\
  $D_s \bar{D}_{s2}^*$ & $2^{--}$ & $4540.9^{+0.4}_{-1.7}$ & $4499^{+13}_{-19}$ & $-$ & ${4533.9^{+3.5}_{-12.7}}$ & ? & ?  \\
  \hline
  $D_s^* \bar{D}_{s1}$ & $0^{-+}$ & $-$ & $-$ & ${(4645.8^{+1.5}_{-10.6})}^V$ & $-$ & ? & ? \\
  $D_s^* \bar{D}_{s1}$ & $0^{--}$ & $4640.7^{+0.5}_{-0.7}$ & $4586^{+20}_{-31}$ & $-$ &
  ${4647.1_{-2.7}^{+0.2}}$ & ? & ? \\
 &  &  &  &  & &  &  \\
  $D_s^* \bar{D}_{s1}$ & $1^{-+}$ & ${(4645.1^{+3.1(B)}_{-11.3})}^V$ & $4583^{+22}_{-32}$ & $-$ & $-$ & $X(4630)$ & $4626^{+24}_{-111}$~\cite{Aaij:2021ivw} \\
 &  &  &  &  & &  &  \\
  $D_s^* \bar{D}_{s1}$ & $1^{--}$ & $4644.2^{+0.1}_{-0.4}$ & $4620.4^{+9.0}_{-13.5}$ & ${(4646.3^{+1.0}_{-14.0})}^V $ & $4643.7_{-6.1}^{+3.2}$ & $Y(4620)$ & $4625.9^{+6.2}_{-6.0}$~\cite{Jia:2019gfe} \\
 &  &  &  &  & &  &  \\
  $D_s^* \bar{D}_{s1}$ & $2^{-+}$ & ${(4648.2^{+0.0}_{-0.9(B)})}^V$ & $4617^{+15}_{-16}$ & ${4647.3^{+8.1}_{-4.7}}$  & $4647.3^{+0.0}_{-4.3}$ & $X(4630)$ & $4626^{+24}_{-111}$~\cite{Aaij:2021ivw} \\
 &  &  &  &  & &  &  \\
  $D_s^* \bar{D}_{s1}$ & $2^{--}$ & $4647.3^{+0.9(V)}_{-0.0}$ &  $4642.6^{+1.9}_{-10.8}$ & ${(4646.0^{+1.2}_{-12.3})}^V$ & $4635.6_{-6.2}^{+5.1}$  & ? & ? \\ 
  \hline
  $D_s^* \bar{D}_{s2}^*$ & $1^{-+}$ & $-$ & $-$ & $-$ & $-$ & ? & ? \\
  $D_s^* \bar{D}_{s2}^*$ & $1^{--}$ & $-$ & $-$ & $-$ & $-$ & ? & ? \\
  $D_s^* \bar{D}_{s2}^*$ & $2^{-+}$ & $-$ & $-$ & $-$ & $-$ & ? & ? \\
  $D_s^* \bar{D}_{s2}^*$ & $2^{--}$ & $-$ & $4662.9^{+6.1}_{-9.2}$ & $-$ & $-$ & ? & ? \\
  $D_s^* \bar{D}_{s2}^*$ & $3^{-+}$ & $-$ & $-$ & $-$ & $-$ & ? & ? \\
  $D_s^* \bar{D}_{s2}^*$ & $3^{--}$ & $4681.7^{+0.4}_{-0.8}$ & $4656.9^{+8.1}_{-12.2}$ & $4640^{+14}_{-21}$ & ${4624^{+19}_{-29}}$ & ? & ? \\
  \hline\hline 
\end{tabular}
\caption{Extension of the spectrum of Table \ref{tab:predictions-nonstrange}
  to the hidden strange sector, i.e. to the $D_{s1(s2)}^{(*)} \bar{D}_s^{(*)}$
  molecules.
  The conventions are identical as in Table \ref{tab:predictions-nonstrange},
  except for the superscript $V$ that denotes a virtual state solution.
}
\label{tab:predictions-strange}
\end{table*}

\section{Conclusions}
\label{sec:conclusions}

In this manuscript we have considered the $Y(4230)$, $Y(4360)$, $Y(4500)$ and
$Y(4620)$ resonances from a molecular perspective, where the interaction
between the charmed meson and antimeson is constrained by their heavy- and
light-quark symmetries, i.e. HQSS and SU(3)-flavor.
Though it is unlikely that the $Y(4230)$ or any other of the aforementioned
resonances is a purely molecular state, it is also not particularly
probable that they are completely non-molecular.
For simplicity we have treated these states as pure molecules,
without admixture from shorter-range charmonium components,
and then built an EFT description of the potential between
their hadronic components, which at leading order only
involves contact-range interactions.

The resulting EFT potential contains four unknown couplings per isospin or
SU(3)-flavor representation, i.e. eight couplings in total.
This implies that there are not enough molecular candidates available
to determine these couplings, a limitation that we propose
to overcome by invoking a series of simplifications.
On the basis of a preexisting light-meson exchange phenomenological
model~\cite{Peng:2021hkr},
we are motivated to make the assumption that the interaction
between the S- and P-wave charmed mesons conforming the $Y(4230)$ state
is dominated by a non-diagonal $D \bar{D}_1 \to D_1 \bar{D}$ term
involving an electric-like dipolar E1 $D \to D_1$ transition
mediated by vector-meson exchange. 
In this case, which we call the $Y$-counting,
a molecular $Y(4230)$ will imply the existence of
a few partners, among which there are states with mass and quantum
numbers potentially compatible with those of the $Y(4360)$,
$Y(4500)$ and $Y(4620)$.
The $Y(4360)$ can be interpreted as a $J^{PC} = 1^{--}$ $D^* \bar{D}_1$
bound state, the $Y(4500)$ as a $J^{PC} = 1^{--}$ $D_s \bar{D}_{s1}$ one
and the $Y(4620)$ as a $J^{PC} = 1^{--}$ $D_s^* \bar{D}_{s1}$ virtual state,
with this later state probably having $2^{--}$ $D_s^* \bar{D}_{s1}$  and
$3^{--}$ $D_s^* \bar{D}_{s2}^*$ partners of
similar mass.

The previous predictions can be further refined by considering a second
non-diagonal $D^* \bar{D}_1 \to D_1 \bar{D}^*$ term involving a magnetic-like
quadrupolar M2 contribution to the $D^* \to D_1$  transition.
We call this possibility the $Y^*$-counting, as we determine the new coupling
associated with the M2 interaction from the mass of the $Y(4360)$.
The inclusion of this interaction improves the predictions of the mass
of the candidate molecular configurations for the $Y(4620)$ ---
the $1^{--}$ $D_s \bar{D}_{s1}$ configuration --- which
now becomes a bound state.

Yet, a further improvement is possible: the mass difference of the ground and
excited states of the P-wave charmed mesons is merely $39\,{\rm MeV}$
($34\,{\rm MeV}$) for the non-strange (strange) case, suggesting
the inclusion of the coupled channel effects
related to the $D_1 \to D_2^*$ transitions.
This choice results for instance in a moderate improvement of the attraction
and binding of the candidate configuration for the $Y(4620)$.
The molecules where coupled channel effects are more conspicuous
are the $J^{PC} = 1^{-+}$ and $2^{-+}$ $D_s^* \bar{D}_{s1}$ systems,
which are both attractive now.
In particular, there is a near-threshold $2^{-+}$ $D_s^* \bar{D}_{s1}$ bound state
that might very well correspond to the $X(4630)$ discovered by
the LHCb~\cite{Aaij:2021ivw},
the spin and parity of which are expected to be $J^P = 1^-$ or $2^-$.

However there are limitations to the conclusions we draw in this manuscript.
For starters neither the $Y(4230)$ or $Y(4360)$ are expected to be fully
molecular, particularly when one considers their relatively
large binding energies, which suggests that the $c\bar{c}$ components
  of their wave functions~\cite{LlanesEstrada:2005hz,Li:2009zu}
  are probably important.
Yet, we believe that the symmetry considerations on which our analysis
rests will more likely than not supersede the specific details
about the structure of these two resonances: the $Y(4500)$ and
$Y(4620)$ are actually predicted as hidden-strange
partners of the $Y(4230)$ or $Y(4360)$.
This point could be elucidated by future theoretical analyses.

There is a second caveat regarding the two power countings built for deriving
our conclusions, which are based on the assumption that the non-diagonal
E1 and M2 interactions are considerably stronger than the central and
spin-spin ones.
Even though there might be reasons to think this is the case, the numerical
factors in front of these E1 and M2 terms change considerably
for the different charmed meson-antimeson configurations,
particularly when the total spin is low.
As a consequence these power countings are expected to work better
in molecules such as $D \bar{D}_1$, $D \bar{D}_2^*$ and $J=3$
$D^* \bar{D}_2^*$ (for which the numerical factor in front of the E1 and M2
couplings --- $D_{0b}$ and $D_{0c}$ in this work --- is close to one),
but not so well for the other configurations of these systems.
This means that it is more likely than not that we are underpredicting
the amount of attraction of a few configurations of the
$D^* \bar{D}_1$ and $D_s^* \bar{D}_{s1}$ systems.
Of particular importance here is the sign of the spin-spin coupling:
an attractive spin-spin term will reinforce the conclusion that
the $X(4630)$ is a $2^{-+}$ $D_s^* \bar{D}_{s1}$ molecule, while
a repulsive one will favor binding in the $1^{-+}$ $D_s^* \bar{D}_{s1}$
configuration, which will thus become the preferred
interpretation of the $X(4630)$ (as happens
in~\cite{Yang:2021sue,Wang:2021ghk}).
Here we have calculated the molecular spectrum
  in two phenomenological models, both of which support the idea of
  binding in the $2^{-+}$ $D_s^* \bar{D}_{s1}$ system, but also showing
  that the $1^{-+}$ $D_s^* \bar{D}_{s1}$ molecule could bind as well.
The eventual experimental observation of molecular candidates in these systems
together with improved lattice calculations (in the line of
Refs.~\cite{Prelovsek:2020eiw,Padmanath:2022cvl}) or
more refined phenomenological models might be used to overcome
the ambiguity of the previous results in the future.

\section*{Acknowledgments}

We would like to thank Xiang-Kun Dong for a careful reading of the manuscript.
This work is partly supported by the National Natural Science Foundation
of China under Grants No. 11735003, No. 11835015, No. 11975041, No. 12047503
and No. 12125507, the Chinese Academy of Sciences under Grant No. XDB34030000,
the China Postdoctoral Science Foundation under Grant No. 2022M713229,
the Fundamental Research Funds for the Central Universities and
the Thousand Talents Plan for Young Professionals.
M.P.V. would also like to thank the IJCLab of Orsay, where part of
this work has been done, for its long-term hospitality.

\appendix

\section{One pion exchange potential}
\label{app:OPE}

Here we derive the form of the one pion exchange (OPE) potential
for the $D_{1(2)}^{(*)} \bar{D}^{(*)}$ molecules.
In particular we are interested in whether OPE is perturbative or not:
previous experience in the $D^{(*)}\bar{D}^{(*)}$ two-body system, where OPE
is indeed perturbative~\cite{Fleming:2007rp,Valderrama:2012jv},
suggests that this is likely to be the case too
when one of the S-wave charmed mesons is substituted
by a P-wave charmed meson.
Yet, owing to the differences between these two systems,
this hypothesis should be explicitly checked.
As we will see, OPE is most probably still perturbative
when P-wave charmed mesons are involved.

First, we begin by noticing that there are two types of OPE
for the $D_{1(2)}^{(*)} \bar{D}^{(*)}$ molecules:
diagonal (e.g $D^* \bar{D}_2^* \to D^* \bar{D}_2^*$)
and non-diagonal (e.g $D^* \bar{D}_2^* \to D_2^* \bar{D}^*$).
For the diagonal one, the relevant Lagrangians
in the superfield notation are
\begin{eqnarray}
  \mathcal{L}_{\pi H H} &=&
  g_1\,{\rm Tr}\left[H^{\dagger}\, \vec{\sigma} \cdot \vec{A}\, H \right] \, , \\
  \mathcal{L}_{\pi T T} &=&
  g_1'\,{\rm Tr}\left[T_k^{\dagger}\, \vec{\sigma} \cdot \vec{A}\, T_k \right] \, ,
\end{eqnarray}
which are basically the non-relativistic limit of the Lagrangians
in~\cite{Falk:1992cx}, and where $g_1$ and $g_1'$ are the axial
coupling constants, $H$ and $T_k$ the S- and P-wave charmed meson
superfields and $\vec{A}$ represents the non-relativistic
axial pion current, which we define as
\begin{eqnarray}
  \vec{A} = \frac{1}{\sqrt{2} f_{\pi}} \vec{\nabla}\,(\tau_a \pi_a) \, ,
\end{eqnarray}
with $f_{\pi} \simeq 132\,{\rm MeV}$ the weak pion decay constant,
$\tau_a$ the Pauli matrices in isospin space, $\pi_a$ the pion field
and $a = 1,2,3$ an isospin index.
If we choose the subfield notation instead, then the Lagrangians will read
\begin{eqnarray}
  \mathcal{L}_{\pi q_s q_s} &=& g_1\,
  q_s^{\dagger}\, \vec{\sigma}_L \cdot \vec{A}\, q_s \, , \\
  \mathcal{L}_{\pi q_p q_p} &=& g_1'\,
  q_p^{\dagger}\, \vec{S}_L \cdot \vec{A}\, q_p \, ,
\end{eqnarray}
with $q_s$, $q_p$, $\vec{\sigma}_L$ and $\vec{S}_L$ having
the same meaning as in Eq.~(\ref{eq:LC}).
There is also the non-diagonal pion interaction, which reads
\begin{eqnarray}
  \mathcal{L}_{\pi H T} &=&
  \frac{h_1}{\Lambda_{\chi}}\,
       {\rm Tr}\left[ H^{\dagger} T_i \partial_i \sigma_j A_j \right]
  +
  \frac{h_2}{\Lambda_{\chi}}\,
       {\rm Tr}\left[ H^{\dagger} T_i \partial_j \sigma_j A_i \right] \nonumber
        \\ &+& {\rm C.C.} \, , \label{eq:OPE-Wc-Lagrangian}
\end{eqnarray}
in the standard superfield notation~\cite{Falk:1992cx}, with $\Lambda_{\chi}$
a typical chiral energy scale (usually chosen to be around $1\,{\rm GeV}$) and
where the difference between the terms proportional to the $h_1$ and
$h_2$ couplings lies in how the subindices of the $T$ superfield
are contracted, i.e. $\vec{T} \cdot \vec{\partial}$ and
$\vec{T} \cdot \vec{A}$ for $h_1$ and $h_2$,
respectively.
If we write this Lagrangian in the subfield notation,
we will obtain instead
\begin{eqnarray}
  \mathcal{L}_{\pi q_s q_p} &=&
  \left[ \frac{h}{\Lambda_{\chi}}\,q_s^{\dagger}\,Q_{L ij}\, q_p - i
    \frac{f}{2 \Lambda_{\chi}}\,q_s^{\dagger}\,\epsilon_{ijk} \Sigma_{L k}
   q_p  \right]\,\partial_i A_j \nonumber \\ &+& {\rm C.C.} \, ,
\end{eqnarray}
where the second term (proportional to the coupling $f$) actually
cancels out (because $\partial_i A_j$ is proportional
to $\partial_i \partial_j \pi_a$ and thus symmetric).
The relation of the couplings $h$ and $f$
with those of Eq.~(\ref{eq:OPE-Wc-Lagrangian}) is
\begin{eqnarray}
  h = h_1 + h_2 \quad \mbox{and} \quad f = h_1 - h_2 \, .
\end{eqnarray}

With the previous Lagrangians we can derive the OPE potential
in momentum space, leading to
\begin{eqnarray}
  V_{\rm OPE}(\vec{q}) &=&
  \frac{g_1 g_1'}{2 f_{\pi}^2}\,\vec{\tau}_1 \cdot \vec{\tau}_2\,
  \frac{\vec{\sigma}_L \cdot \vec{q} \, \vec{S}_L \cdot \vec{q}}{{\vec{q}\,}^2 + m_{\pi}^2} \nonumber \\
  &+&
  \frac{1}{2 f_{\pi}^2}\,\frac{h^2}{\Lambda_{\chi}^2}\,
  \vec{\tau}_1 \cdot \vec{\tau}_2\,
  \frac{Q^{\dagger}_{L ij} Q_{L kl}\,q_i q_j q_l q_k}
       {{\vec{q}\,}^2 + \mu_{\pi}^2} \, ,
\end{eqnarray}
where $\vec{q}$ refers to the exchanged momentum, $m_{\pi} \simeq 138\,{\rm MeV}$
the pion mass and $\mu_{\pi}$ the effective pion mass for the non-diagonal
transitions ($\mu_{\pi}^2 = m_{\pi}^2 - \omega_{\pi}^2$, with
$\omega_{\pi} = m(D_{1(2)}^*) - m(D^{(*)})$).
This potential is not purely S-wave, but contains tensor forces with angular
momenta of $L=2$ for the diagonal piece and $L=2,4$
for the non-diagonal one.
These tensor forces require S-to-D- and S-to-G-wave mixing to operate.
Here we will assume that the higher partial wave components of the molecular
wave functions are kinematically suppressed and consequently
ignore these tensor forces.
The S-wave, finite-range piece of the OPE potential can be easily isolated
and reads
\begin{eqnarray}
  V_{\rm OPE}(\vec{q}) &=&
  -\frac{g_1 g_1'}{6 f_{\pi}^2}\,\vec{\tau}_1 \cdot \vec{\tau}_2\,
  \frac{\vec{\sigma}_L \cdot \vec{S}_L m_{\pi}^2}{{\vec{q}\,}^2 + m_{\pi}^2} \nonumber \\
  &+&
  \frac{1}{2 f_{\pi}^2}\,\frac{2}{15}\,\frac{h^2}{\Lambda_{\chi}^2}\,
  \vec{\tau}_1 \cdot \vec{\tau}_2\,
  \frac{Q^{\dagger}_L \cdot Q_{L}\,\mu_{\pi}^4}
       {{\vec{q}\,}^2 + \mu_{\pi}^2} \nonumber \\
       &+& \dots
\end{eqnarray}
where the dots indicate contact-range contributions or tensor forces.

For determining the momentum scale above which OPE becomes non-perturbative
for S-wave scattering, which we will call $\Lambda_{S}$,
we will rewrite the previous contributions as
\begin{eqnarray}
  V_{\rm OPE}(\vec{q}) &=& \frac{4 \pi}{\mu\,\Lambda_{\rm OPE}}\,
  \vec{\tau}_1 \cdot \vec{\tau}_2\,\hat{\mathcal{O}}_{L12}\,
  \frac{m_{\pi}^2}{{\vec{q}\,}^2 + m_{\pi}^2} + \dots \, ,
  \nonumber \\
\end{eqnarray}
where $\mu$ is the reduced mass of the system,
$\hat{\mathcal{O}}_{L12}$ refers to the specific spin-spin operator
that applies and $\Lambda_{\rm OPE}$ is the characteristic OPE momentum
scale (which is proportional to $\Lambda_S$,
the scale we are looking for).
In this form, the ratio between iterated and tree-level OPE is given by
\begin{eqnarray}
  \frac{\langle V_{\rm S}\,G_0\,V_{\rm S} \rangle}{\langle V_{\rm S} \rangle} =
  \frac{Q}{\Lambda_{S}} = T S \frac{m_{\pi}}{\Lambda_{\rm OPE}}\,
  f(\frac{k}{m_{\pi}}) \, , \label{eq:VGV-ratio}
\end{eqnarray}
where the operators have been evaluated for an S-wave scattering state,
$V_S$ is the S-wave projection of the OPE potential, $G_0(E) = 1/(E-H_0)$
is the resolvent operator (with $E$ is the center-of-mass energy and
$H_0$ the free Hamiltonian), $T$ ($S$) refer to the evaluation of
the $\tau_1 \cdot \tau_2$ ($\hat{\mathcal{O}}_{L12}$) operator and
$f(x) = 1 - 13/6 x^2 + \mathcal{O}(x^4)$ is a function
that encodes the momentum dependence of this ratio (details on
how this function is calculated are to be found
in Ref.~\cite{PavonValderrama:2016lqn}).
For simplicity we will take $f(x) = 1$ from now on.

The bottom-line is that for center-of-mass momenta $k < \Lambda_S$,
the previous ratio will be smaller than $1$ and OPE will behave perturbatively.
The estimations we obtain for $\Lambda_S$ are the following
\begin{eqnarray}
  \Lambda_S^{({\rm D})} &=& \frac{4 \pi}{\mu}\,6\,f_{\pi}^2\,\frac{1}{g_1 g_1'}\,\
  \frac{1}{T S} \, , \\
  \Lambda_S^{({\rm ND})} &=& \frac{4 \pi}{\mu}\,15\,f_{\pi}^2\,
  \frac{\Lambda_{\chi^2}}{\mu_{\pi}^2}\,\frac{1}{h^2}\,\frac{1}{T S} \, ,
\end{eqnarray}
for the diagonal and non-diagonal pieces, respectively.
The only thing that remains for the determination of $\Lambda_S$ is
the value of the axial couplings, $g_1$, $g_1'$ and $h$.
For the first, we originally had the determination
$g_1= 0.59 \pm 0.01 \pm 0.07$~\cite{Ahmed:2001xc,Anastassov:2001cw}
from the strong $D^*$ decays, which can be refined further by using
the updated $D^*$ decay widths~\cite{Zyla:2020zbs},
leading to the estimations $g_1 = 0.54$~\cite{Mehen:2015efa} or
$g_1 = 0.56 \pm 0.01$~\cite{Yan:2021wdl}).
For the second, $g_1'$ is not directly extractable in terms of decays
($D_2^* \to D_1 \pi$ is not energetically possible) but
it can be estimated from the quark model
leading to $g_1' = (2/3)\,g_1$.
For the last, $h$ appears in the strong P-wave meson decays, e.g.
\begin{eqnarray}
  \Gamma(D_1 \to D^*\pi) = \frac{1}{4\pi}\,\frac{m_{D^*}}{m_{D_1}}\,
  \frac{h^2}{\Lambda_{\chi}^2}\,\frac{q_{\pi}^5}{f_{\pi}^2} \, ,
\end{eqnarray}
and the analogous expressions for the other two decays
($D_2^* \to D \pi$ and $D_2^* \to D^* \pi$).
The $D_1 \to D^*\pi$ decay yields $h \simeq 0.9$, where this value might
be an overestimation owing to the existence of some mixing between
the $D_1$ ($J_L = 3/2$) and $D_1^*$ ($J_L=1/2$) charmed mesons,
meaning that not all of the $D_1$ strong decay width
is derived from the $h$ coupling.

With the previous numbers we obtain that
\begin{eqnarray}
  \Lambda_S^{({\rm D})} \simeq \frac{5.9\,{\rm GeV}}{T S}
  \quad \mbox{and} \quad
  \Lambda_S^{({\rm ND})} \simeq \frac{190\,{\rm GeV}}{T S} \, ,
\end{eqnarray}
which are relatively large, particularly for the non-diagonal component.
The lowest value of the $\Lambda_S$ scales happens for $T = 3$ and $S=5/2$
(both for the diagonal and non-diagonal cases), bringing down
the previous numbers to $\Lambda_S^{({\rm D})} = 791\,{\rm MeV}$ and
$\Lambda_S^{({\rm ND})} = 25.9\,{\rm GeV}$.
Yet, even in this case, the diagonal S-wave OPE interaction only becomes
non-perturbative at momenta of the order of the $\rho$ mass at least,
and most probably at higher momenta owing to the neglected
$f(k/m_{\pi})$ factor in Eq.~(\ref{eq:VGV-ratio}), which
works towards making pions more perturbative.
The perturbativeness of non-diagonal OPE is more extreme, where the reason
is in part the larger numerical factor in from of $f_{\pi}^2$
for the non-diagonal piece as well as the dimensional
factor $\Lambda_{\chi}^2 / \mu_{\pi}^2$.
Basically, this reflects the higher angular momentum and dimensionality
of the operators involved in the $D^{(*)} \to D_{1(2)}^{(*)}$ transitions,
where the contribution that survive in S-wave scattering is really small.

\section{Transition spin operators}
\label{app:transition-operators}

The components of the light-spin operators $\vec{\Sigma}_{L}$
for the spin-$\tfrac{1}{2}$ to spin-$\tfrac{3}{2}$
transition are
\begin{eqnarray}
  \Sigma_{L1} &=&
  \begin{pmatrix}
    \frac{1}{\sqrt{2}} & 0 & -\frac{1}{\sqrt{6}} & 0 \\
    0 & \frac{1}{\sqrt{6}} & 0 & -\frac{1}{\sqrt{2}} 
  \end{pmatrix}
  \, , \\
  \Sigma_{L2} &=&
  \begin{pmatrix}
    \frac{i}{\sqrt{2}} & 0 & \frac{i}{\sqrt{6}} & 0 \\
    0 & \frac{i}{\sqrt{6}} & 0 & \frac{i}{\sqrt{2}}
  \end{pmatrix}
  \, , \\
  \Sigma_{L2} &=&
  \begin{pmatrix}
    0 & -\sqrt{\frac{2}{3}} & 0 & 0 \\
    0 & 0 & -\sqrt{\frac{2}{3}} & 0 
  \end{pmatrix}
  \, .
\end{eqnarray}
They are normalized such that
\begin{eqnarray}
  \Sigma_{Li} \Sigma_{Lj}^{\dagger} =
  \frac{2 \delta_{ij} - i \epsilon_{ijk} \sigma_k}{3} \, ,
\end{eqnarray}
where $\sigma_k$ are the Pauli matrices.

Next we consider the matrix elements of the $\vec{\Sigma}_L$ operator
between the S- and P-wave charmed meson states $D^*$ and $D_2^*$.
This results in the definition of a transition matrix $\vec{\Sigma}$ between
spin-1 and spin-2 states, the components of which read:
\begin{eqnarray}
  \Sigma_{1} &=&
  \begin{pmatrix}
    \frac{1}{\sqrt{2}} & 0 & - \frac{1}{2 \sqrt{3}} & 0 & 0 \\
    0 & \frac{1}{2} & 0 & - \frac{1}{2} & 0 \\
    0 & 0 & \frac{1}{2 \sqrt{3}} & 0 & -\frac{1}{\sqrt{2}}  \\
  \end{pmatrix}
  \, , \\
  \Sigma_{2} &=&
  \begin{pmatrix}
    \frac{i}{\sqrt{2}} & 0 & \frac{i}{2 \sqrt{3}} & 0 & 0 \\
    0 & \frac{i}{2} & 0 & \frac{i}{2} & 0 \\
    0 & 0 & \frac{i}{2 \sqrt{3}} & 0 & \frac{i}{\sqrt{2}}  \\
  \end{pmatrix}
  \, , \\
  \Sigma_{3} &=&
  \begin{pmatrix}
    0 & -\frac{1}{\sqrt{2}} & 0 & 0 & 0 \\
    0 & 0 & -\sqrt{\frac{2}{3}} & 0 & 0 \\
    0 & 0 & 0 & -\frac{1}{\sqrt{2}} & 0 \\
  \end{pmatrix}
  \, .
\end{eqnarray}

The evaluation of the product of two $\vec{\Sigma}_L$ operators is
\begin{eqnarray}
  \vec{\Sigma}_{L}^{\dagger} \cdot \vec{\Sigma}_{L} =
  \begin{cases}
    \frac{1}{3} & \quad \mbox{for $S_L=1$} \, , \\
    1 & \quad \mbox{for $S_L=2$} \, ,
  \end{cases} 
\end{eqnarray}
where we have written the result in terms of
the total light-spin of the system.
From this and the heavy- and light-spin decomposition of the different
configurations of the two heavy meson states (which can be calculated
by making use of the 9J symbol formalism described for instance
in Refs.~\cite{Ohkoda:2012rj,Xiao:2013yca,Lu:2017dvm}),
it is possible to extract the matrix elements of
$\vec{\Sigma}_{L}^{\dagger} \cdot \vec{\Sigma}_{L}$
that appear in Table \ref{tab:molecules}
for the $D_b$ coupling.

Alternatively, we might use instead Eqs.~(\ref{eq:SigmaL-1}-\ref{eq:SigmaL-3})
and then calculate directly the corresponding operators for the particular
two charmed meson state under consideration.
In this second case we obtain
\begin{eqnarray}
  {\vec{\epsilon}\,}^* \cdot \vec{\epsilon} &=& 1 \quad \mbox{for $J=1$} \, , \\
  \nonumber \\
  \vec{\Sigma}^{\dagger} \cdot \vec{\Sigma} &=&
  \begin{cases}
    \frac{1}{6} & \quad \mbox{for $J=1$} \, , \\
    \frac{1}{2} & \quad \mbox{for $J=2$} \, , \\
    1 & \quad \mbox{for $J=3$} \, .
  \end{cases} 
\end{eqnarray}

Finally, we consider the quadrupolar operators and their scalar products,
the evaluation of which yields (in terms of light spin)
\begin{eqnarray}
  Q_{L}^{\dagger} \cdot Q_{L} =
  \begin{cases}
    -\frac{5}{2} & \quad \mbox{for $S_L=1$} \, , \\
    +\frac{1}{2} & \quad \mbox{for $S_L=2$} \, ,
  \end{cases} 
\end{eqnarray}
from which, as previously mentioned for
$\vec{\Sigma}_L^{\dagger} \cdot \vec{\Sigma}_L$,
one can derive the matrix elements in the charmed meson basis provided
we have the heavy- and light-spin decomposition of the two-body states.
If we use instead Eqs.~(\ref{eq:QL-1}-\ref{eq:QL-4}),
we can directly write the matrix elements of the projection
of the quadrupolar operators in the charmed meson basis as follows
\begin{eqnarray}
  \epsilon_t^{\dagger} \cdot \epsilon_t &=& +1 \quad \mbox{for $J=2$} \, , \\
  \nonumber \\
  Q^{\dagger} \cdot Q &=&
  \begin{cases}
    +\frac{5}{3} & \quad \mbox{for $J=0$} \, , \\
    -\frac{5}{6} & \quad \mbox{for $J=1$} \, , \\
    +\frac{1}{6} & \quad \mbox{for $J=2$} \, , \\ 
  \end{cases} \\
  \nonumber \\
  Q_{12}^{\dagger} \cdot Q_{12} &=&
  \begin{cases}
    +\frac{3}{4} & \quad \mbox{for $J=1$} \, , \\
    -\frac{5}{4} & \quad \mbox{for $J=2$} \, , \\
    +\frac{1}{2} & \quad \mbox{for $J=3$} \, , \\ 
  \end{cases} 
\end{eqnarray}
from which we will be able to derive the numerical factors found
in Table \ref{tab:molecules} for the $D_c$ coupling.

Finally, we consider coupled-channel effects for the $J=1,2$ molecules
in the bases
\begin{eqnarray}
  \mathcal{B}(J=1) &=&
  \left\{ D\bar{D}_1, D^* \bar{D}_1, D^* \bar{D}_2^*\right\}
  \, , \\
  \mathcal{B}(J=2) &=&
  \left\{ D\bar{D}_2^*, D^* \bar{D}_1, D^* \bar{D}_2^*\right\}
  \, ,
\end{eqnarray}
where the correct C-parity combinations are implicitly assumed
(though not written down).
In the previous bases, the matrix elements of the non-diagonal operators read
  \begin{eqnarray}
    \vec{\Sigma}_L^{\dagger} \cdot \vec{\Sigma}_L
    &=&
  \begin{cases}
    \begin{pmatrix}
      \frac{2}{3} & -\frac{\sqrt{2}}{3} & 0 \\
      -\frac{\sqrt{2}}{3} & \frac{1}{6} & \frac{\sqrt{5}}{6} \\
      0 & \frac{\sqrt{5}}{6} & \frac{1}{6}
    \end{pmatrix} & \quad \mbox{for $J^{PC}=1^{-+}$} \, , \\
    \begin{pmatrix}
      0 & \sqrt{\frac{2}{3}} & 0 \\
      \sqrt{\frac{2}{3}} & \frac{1}{6} & -\frac{1}{2} \\
      0 & -\frac{1}{2} & -\frac{1}{2}
    \end{pmatrix} & \quad \mbox{for $J^{PC}=2^{-+}$} \, , \\ 
  \end{cases}  \\
  \nonumber \\
  {Q}_L^{\dagger} \cdot {Q}_L
    &=&
  \begin{cases}
    \begin{pmatrix}
      0 & 0 & -\sqrt{\frac{5}{2}} \\
      0 & \frac{5}{4} & -\frac{3 \sqrt{5}}{4} \\
      -\sqrt{\frac{5}{2}} & -\frac{3 \sqrt{5}}{4} & -\frac{3}{4}
    \end{pmatrix} & \quad \mbox{for $J^{PC}=1^{-+}$} \, , \\
    \begin{pmatrix}
      1 & 0 & \sqrt{\frac{3}{2}} \\
      0 & \frac{1}{4} & -\frac{3}{4} \\
      \sqrt{\frac{3}{2}} & -\frac{3}{4} & +\frac{5}{4}
    \end{pmatrix} & \quad \mbox{for $J^{PC}=2^{-+}$} \, , \\ 
  \end{cases} \, 
  \end{eqnarray}
  all of which change sign if the C-parity is changed.
  For completeness, though it is not strictly a transition operator,
  it is useful to provide too the matrix elements of
  the spin-spin operator in the previous two bases:
  \begin{eqnarray}
    \vec{\sigma}_L \cdot \vec{S}_L
    &=&
  \begin{cases}
    \begin{pmatrix}
      0 & -\frac{5}{2\sqrt{2}} & -\frac{1}{2}\sqrt{\frac{5}{2}} \\
      -\frac{5}{2\sqrt{2}} & -\frac{5}{4} & \frac{\sqrt{5}}{4} \\
      -\frac{1}{2}\sqrt{\frac{5}{2}} & \frac{\sqrt{5}}{4} & -\frac{9}{4}
    \end{pmatrix} & \quad \mbox{for $J^{P}=1^{-}$} \, , \\
    \begin{pmatrix}
      0 & \frac{1}{2}\sqrt{\frac{3}{2}} & -\frac{3}{2}\sqrt{\frac{3}{2}} \\
      \frac{1}{2}\sqrt{\frac{3}{2}} & \frac{5}{4} & \frac{3}{4} \\
      -\frac{3}{2}\sqrt{\frac{3}{2}} & \frac{3}{4} & -\frac{3}{4}
    \end{pmatrix} & \quad \mbox{for $J^{P}=2^{-}$} \, , \\ 
  \end{cases}  
  \end{eqnarray}
  which are independent of C-parity, as these are direct terms.

\section{Phenomenological approaches}
\label{app:phenomenology}

Here we explain the two phenomenological models we used to make the
predictions of S- and P-wave charmed meson-antimeson molecules
in Tables \ref{tab:predictions-nonstrange} and
\ref{tab:predictions-strange}, namely
the OBE and the RG-saturation models.

\subsection{The one boson exchange model}

First, we consider the time-honored OBE model and how to calculate
the molecular spectrum in it.
Within the OBE model,
the potential between two hadrons is constructed as the sum of
the potentials generated by the exchange of a few light mesons,
with the most common choice being the pseudoscalar (the pion),
scalar (the sigma) and vector (the rho and the omega) mesons.
Owing to the expected perturbativeness of OPE
(discussed in Appendix~\ref{app:OPE}) we will ignore the pion from now on.
We already presented the potentials generated by scalar and vector meson
exchanges in Eqs.~(\ref{eq:Va}-\ref{eq:Wc}), yet within the OBE model
these potentials are modified as a consequence of the finite size of
the hadrons involved.
For this, a form factor and a cutoff are included as follows:
\begin{eqnarray}
  V_{\rm M(finite)}(\vec{q}\, , \Lambda_{\rm OBE}) = V_{\rm M(point-like)}(\vec{q}\,)\,
  f^2_M(\vec{q}) \, ,
\end{eqnarray}
with $V_{\rm M(point-like)}$ the potential for point-like hadrons,
i.e. Eqs.~(\ref{eq:Va}-\ref{eq:Wc}), $V_{\rm M(finite)}$ the potential
for finite-size hadrons and where $f_M(x)$ is the form factor
representing the finite-size effects
for the exchange of meson $M$.
In particular we will consider multipolar form factors of the type
\begin{eqnarray}
  f_M(\vec{q}) =
  {\left( \frac{\Lambda_M^2 - m^2}{\Lambda_M^2 - {q}^2} \right)}^{2 n_F} \, ,
\end{eqnarray}
where $m$ is the mass of the exchanged meson, $q^2 = -q_0^2 + {\vec{q}\,}^2$
the square of the 4-momentum, $\Lambda_M$ a cutoff and $n_F$
and exponent.
In principle it is possible to use a different form factor and cutoff
for every of the exchanged mesons, which is what is done
when the OBE is applied in the two-nucleon sector~\cite{Machleidt:1987hj,Machleidt:1989tm}.
But here we will limit ourselves to a unique form factor and cutoff (otherwise
there will be not enough input data to determine the cutoffs).

Once the cutoff and the exponent have been chosen predictions are easy
to make by solving the resulting potential in the Schr\"odinger equation.
For the exponent we will choose $n_F = 3$, while the cutoff is chosen
by applying the OBE model to the $X(3872)$ as a $1^{++}$ $D^* \bar{D}$
molecule and then determining the cutoff from the condition of
reproducing the mass of the $X(3872)$.
This will lead to $\Lambda_{\rm OBE} = 1.56\,{\rm GeV}$
for the parameters we specify below.

Besides the form-factor and the cutoff, the parameters of the OBE model are
the masses and couplings of the light mesons that are exchanged.
For the masses we will use $m_{\sigma} = 475\,{\rm MeV}$,
$m_\rho = 770 \rm MeV$ and $m_{\omega} = 780\,{\rm MeV}$.
For the couplings we begin with the ones appearing in the
diagonal components of the potential, Eqs.~(\ref{eq:Va}) and (\ref{eq:Vb}),
for which we take  $g_{\sigma} = 3.4$ (derived from the linear sigma
model~\cite{GellMann:1960np} and the quark model~\cite{Riska:2000gd}),
$g_V = 2.9$ (from the universality of the vector
meson coupling~\cite{Sakurai:1960ju,Kawarabayashi:1966kd,Riazuddin:1966sw},
which requires $g_V = m_{V} / \sqrt{2} f_{\pi}$
with $f_{\pi} = 132\,{\rm MeV}$), $f_{V} = g_V \kappa_{Vi}$ 
with $\kappa(D^{(*)}) = 2.9$ and
$\kappa(D_{1(2)}^{(*)}) = 2 \kappa(D^{(*)})$~\cite{Peng:2021hkr}.
For the non-diagonal components of the potential, Eqs.~(\ref{eq:Wb}) and
(\ref{eq:Wc}), we define $f_V' = g_V \kappa_{E1}$ and $h_V' = g_V \kappa_{M2}$ and
determine $\kappa_{E1}$ and $\kappa_{M2}$ from the condition of reproducing
the masses of the $Y(4230)$ and $Y(4360)$ as $1^{--}$ $D \bar{D}_1$
and $D^* \bar{D}_1$ molecules.
This results in $\kappa_{E1} \simeq 4.9$ and $\kappa_{M2} \simeq 22.8$.

\subsection{The RG-improved saturation model}

The second phenomenological model we will consider is that of a contact-range
theory in which the strength of the couplings constants is derived from
the condition that they are saturated by light-meson exchanges~\cite{Ecker:1988te,Epelbaum:2001fm}.
The idea is that at the low momenta characteristic of a hadronic bound state
the details of the light-meson exchange potential are not resolved and
thus its effects can be reduced to a contact-range interaction of
suitable strength.
For a generic light-meson exchange potential of the type
\begin{eqnarray}
  V_M(\vec{q}) = \frac{v_M}{m^2 + {\vec{q}\,}^2} + \dots \, , \label{eq:V-M}
\end{eqnarray}
where the subindex $M$ indicates the particular meson considered,
$v_M$ is a prefactor containing coupling constants and operators and
the dots denote either Dirac-deltas or higher angular momentum components,
the equivalent saturated coupling will be proportional to
\begin{eqnarray}
  C_{M}(\Lambda_{\rm sat} \sim m) \propto \frac{v_M}{m^2} \, , \label{eq:sat-M}
\end{eqnarray}
where $\Lambda_{\rm sat}$ refers to the regularization scale used to regularize
the contact-range potential, which is expected to be of the order of
the mass of the exchanged light-meson.
The proportionality constant is in principle unknown.
It is also important to stress that the previous refers to the specific
saturation method proposed in~\cite{Peng:2020xrf},
which explicitly ignored the contact-range pieces of the light-meson
exchange potentials --- the dots in Eq.~(\ref{eq:V-M}) --- as they
represent physics at scales corresponding to the hadron internal
structure (instead of the light-meson exchange range,
which is what interest us).

When considering the saturation from the scalar and vector mesons, there is
the problem of how to combine them properly: the mass of the scalar and
vector mesons, though of the same order of magnitude, are different
enough as to justify different values of $\Lambda_{\rm sat}$.
What we will do instead is to combine their contributions
by means of the RG equations proposed
in~\cite{PavonValderrama:2014zeq}, which in practical
terms implies the following rule for combining the saturation
from the scalar and vector mesons
\begin{eqnarray}
  C^{\rm sat}(m_V) = C_V(m_V) +
   {\left( \frac{m_V}{m_S} \right) }^{\alpha} \,
    C_{S}(m_S) \, ,
\end{eqnarray}
with $m_S$ and $m_V$ the scalar and vector meson masses, respectively, and
$\alpha$ an exponent related with the behavior of the wave function at
the distances $m_V r = \mathcal{O}(1)$ and $m_S r = \mathcal{O}(1)$.
For a detailed explanation we refer to~\cite{Peng:2021hkr},
where it was also proposed the value $\alpha = 1$
for the exponent.

With all this, the saturated contact-range coupling
for the $D^{(*)} \bar{D}_{1(2)}^{(*)}$ molecules is
given (modulo a proportionality constant) by
\begin{eqnarray}
  && C^{\rm sat}(m_V) \propto
  - {(\frac{m_V}{m_S})}^{\alpha}\frac{g_{S1} g_{S2}}{m_S^2} \nonumber \\ 
  && \quad + \left( \vec{\tau}_1 \cdot \vec{\tau}_2 + \zeta  \right)\,
  g_{V1} g_{V2}\,\Big[ 
    1 +
    \frac{2}{3}\frac{\kappa_{V1} \kappa_{V2}}{6 M^2}\,
    \vec{\sigma}_{L} \cdot \vec{S}_L
    \nonumber \\
  && \quad - 
  \frac{\kappa_{E1}^2}{4 M^2}\,
       {\left( \frac{m_V}{\mu_V} \right)}^{\alpha}\,\left( \frac{\omega_V^2 + \frac{1}{3}\mu_V^2}{\mu_V^2} \right)\,\vec{\Sigma}_L^{\dagger} \cdot \vec{\Sigma}_L
       \nonumber \\
       && \quad - 
       \frac{\kappa_{M2}^2}{16 M^4}\,
            {\left( \frac{m_V}{\mu_V} \right)}^{\alpha}\,\frac{\mu_V^2}{5}\,
            Q_L^{\dagger} \cdot Q_L \Big] \, , 
    \label{eq:coupling-sat}
\end{eqnarray}
where $\kappa_V = f_V / g_V$, $\kappa_{E1} = f_V' / g_V$,
$\kappa_{M2} = h_V' / g_V$ and with $m_S$, $m_V$, $\mu_V$, $\omega_V$ and $M$
as defined in Eqs.~(\ref{eq:Va}-\ref{eq:Wc}).
The proportionality constant can be determined from the condition of
reproducing the mass of a given molecular state.
For this, the saturated contact-range coupling is to be regularized as usual
\begin{eqnarray}
  \langle k | V^{\rm sat}_C(\Lambda_{\rm sat}) | p \rangle =
  C^{\rm sat}(\Lambda_{\rm sat})\,f(\frac{k}{\Lambda_{\rm sat}})\,
  f(\frac{p}{\Lambda_{\rm sat}})\, , \nonumber \\
\end{eqnarray}
where $\Lambda_{\rm sat} \sim m_V$ is the saturation cutoff and $f(x)$
a regulator function.
Then we plug this potential into the bound state equation, Eq.~(\ref{eq:BSE}),
in order to determine the masses of the molecular states.
For concreteness we take $\Lambda_{\rm sat} = 1.0\,{\rm GeV}$, $f(x) = e^{-x^2}$
and use the $X(3872)$ as a $1^{++}$ $D^*\bar{D}$ bound state as the reference
state from which to fix the proportionality constant.
For the masses and couplings we use the same values as for the OBE model except
for $\kappa_{E1}$ and $\kappa_{M2}$, which we determine again from reproducing
$Y(4230)$ and $Y(4360)$ as $1^{--}$ $D \bar{D}_1$ and $D^* \bar{D}_1$
bound states, resulting in $\kappa_{E1} \simeq 3.9$ and
$\kappa_{M2} \simeq 24.5$.

%

\end{document}